\documentclass[%
 reprint,
superscriptaddress,
 amsmath,amssymb,
 aps,
prc,
]{revtex4-2}
\usepackage{graphicx}
\usepackage{dcolumn}
\usepackage{bm}
\usepackage{color}

\usepackage{multirow}
\usepackage{mathtools}

\begin{document}
\title
{ Detectability of a phase transition in neutron star matter with third-generation gravitational wave interferometers }

\author{C. Mondal}
  \email{chiranjib.mondal@ulb.be}
  \affiliation{Normandie Univ., ENSICAEN, UNICAEN, CNRS/IN2P3, LPC Caen, 14000 Caen, France}
  \affiliation{Institut d’Astronomie et d’Astrophysique, Universit\'e Libre de Bruxelles, CP 226, B-1050 Brussels, Belgium}
  \author{M. Antonelli}
  \email{antonelli@lpccaen.in2p3.fr}
  \affiliation{Normandie Univ., ENSICAEN, UNICAEN, CNRS/IN2P3, LPC Caen, 14000 Caen, France}
  \author{F. Gulminelli}
  \email{gulminelli@lpccaen.in2p3.fr}
  \affiliation{Normandie Univ., ENSICAEN, UNICAEN, CNRS/IN2P3, LPC Caen, 14000 Caen, France}
  \author{M. Mancini}
  \email{marco.mancini@univ-orleans.fr}
  \affiliation{IDP, UMR 7013 - CNRS - Univ. Orléans- Univ. Tours.
Université d'Orléans, rue de Chartres, BP 6759,  45067 Orléans Cedex 2, France}
  \affiliation{Laboratoire Univers et Th\'eories, Observatoire de Paris,
 Universit\'e PSL, CNRS, 92190 Meudon, France}
  \author{J. Novak}
  \email{jerome.novak@obspm.fr}
  \affiliation{Laboratoire Univers et Th\'eories, Observatoire de Paris,
 Universit\'e PSL, CNRS, 92190 Meudon, France}
  \author{M. Oertel}
  \email{micaela.oertel@obspm.fr} 
  \affiliation{Laboratoire Univers et Th\'eories, Observatoire de Paris,
 Universit\'e PSL, CNRS, 92190 Meudon, France}






\begin{abstract}
Possible strong first-order hadron-quark phase transitions in neutron star interiors leave an imprint on gravitational waves, which could be detected with planned third-generation interferometers. 
Given a signal from the late inspiral of a  
binary neutron star (BNS) coalescence, 
assessing the presence of such a phase transition depends on the precision that 
can be attained in the determination of the tidal deformability parameter, as well as on the model used to describe the hybrid star equation of state. For the latter, we employ here a phenomenological meta-modelling of the equation of state that largely spans the parameter space associated with both the low density phase and the quark high density compatible with current constraints. 
We show that with a network of third-generation detectors, a single loud BNS event might be sufficient to infer the presence of a phase transition at low baryon densities with an average Bayes factor $B\approx 100$, up to a luminosity distance ($\mathcal{D}_L \lesssim$ 300 Mpc).
\end{abstract}

\maketitle


\section{Introduction}
\label{sec:intro}

In the standard picture, the interior of a neutron star (NS) encompasses several phases of matter, from nuclei in the iron region close to the
surface via more exotic neutron-rich nuclear clusters in the crust to
a uniform liquid of nuclear matter in the outer core \cite{haensel_book}.  
In the core, at densities above nuclear saturation, $n_0 \approx 0.16$
fm$^{-3}$, the pressure becomes so high that further degrees
of freedom may emerge. These include, possibly, mesons --which could
also form condensates--, hyperons or $\Delta$-baryons.  An even more
dramatic possibility is that of a phase transition to degenerate quark
matter and the formation of so-called hybrid stars. Under the
hypothesis of absolutely stable strange quark matter, even pure quark
stars might exist, see e.g. the reviews \cite{FiorellaBurgio:2018dga,Oertel:2016bki,Lattimer:2019iye,Chamel:2008ca,Raduta:2022elz}. This makes the compact objects a unique testbed of subatomic physics, not only probing nuclear many-body phenomena and their dependence on density and isospin asymmetry,
i.e. the neutron-to-proton ratio but also probing unexplored
finite-density regimes of quantum chromodynamics. 

Currently, the main astrophysical constraints on NS interiors stem from the precise mass determinations for three NSs in NS-white dwarf
systems~\cite{Demorest:2010bx,Antoniadis:2013pzd,Fonseca:2016tux,Cromartie:2019kug}, all three with gravitational masses around $2\ M_\odot$. 
The object PSR J0740+6620~\cite{Cromartie:2019kug} is thereby particularly interesting since NICER succeeded in obtaining a measurement of its
radius~\cite{Miller:2021qha,Riley:2021pdl}, making it the second object after
PSR J0030+0451 \cite{Miller:2019cac,Riley:2019yda} for which both mass and radius could be determined. 
Moreover, with the detections of gravitational waves from binary neutron star
(BNS) mergers by the LIGO-Virgo collaboration, a new window has opened
to explore the constituents of matter under extreme conditions. 
For the first detected event, GW170817, the tidal deformability was
obtained~\cite{LIGOScientific:2017vwq}, a quantity strongly correlated with the NS
radius. In the coming years, starting with run O4 of the
LIGO-Virgo-Kagra (LVK) collaboration, the gravitational wave (GW) detector network
sensitivity will be further increased and a number of additional detections is expected. 
Projects for ground based third-generation detectors such as the European Einstein Telescope~\cite{Punturo:2010zz,Maggiore:2019uih} and the American Cosmic Explorer~\cite{Reitze:2019iox,Evans:2021gyd} planned for $\sim$2035 will allow for a considerable gain in sensitivity. They need
more stringent constraints on theoretical models for the description of the dense matter. 
  
A particular question in this context is the possible presence of a first-order
phase transition (PT) in dense neutron star matter. Since the first
mention of possible hybrid stars decades ago, e.g. \cite{Glendenning:1991ic},
finding astrophysical signatures of such a PT has been a very active
field of research, see for instance the reviews
\cite{Alford:2001dt,Buballa:2014jta,Blaschke_new}. In this work, we investigate if and how
accurately we can detect a PT in the core of two coalescing neutron stars
during the inspiral phase with a network of third-generation
gravitational wave detectors. Several authors have already pointed out
that a PT in the post-merger phase leads to a characteristic increase
in the dominant post-merger oscillation
frequency~\cite{Bauswein:2018bma,Most:2018eaw} with respect to the one expected
from the measured inspiral parameters under the assumption of a purely
baryonic equation of state (EoS). If the post-merger signal is
detected it could even help to constrain the onset density for the
phase transition~\cite{Blacker:2020nlq} and a delayed PT could leave an
imprint on the ringdown of the black hole formed once the metastable
remnant has collapsed~\cite{Weih:2019xvw}. These ideas apply if, prior to merger, both stars
do not present any PT, which is plausible for not too
massive stars.

The horizon for detecting a post-merger signal is, however, relatively
small even for third-generation detectors and only a few events are
expected, see e.g.~\cite{Torres-Rivas:2018svp}. On the contrary, a huge
number of BNS mergers should be detected with information extracted
from the inspiral among others about the chirp mass $\mathcal{M}_c$, the
mass ratio $q$ and on the combined tidal deformability
$\tilde\Lambda$ (see Eq.~\eqref{e:def_lambda}), which is a function of the tidal deformabilities
$\Lambda_1$ and $\Lambda_2$ of both stars~\cite{Maggiore:2019uih,Evans:2021gyd,Branchesi:2023mws}. Since a
PT changes the relation between the mass and the tidal deformability
for each star, see e.g.~\cite{Damour:2009vw,Postnikov:2010yn,Sieniawska:2018zzj,Han:2018mtj}, $\tilde\Lambda$ is modified, and this of course raises the question of its detectability which many studies have addressed recently.

For instance, based on the breakdown of quasi-universal relations
fitted to purely hadronic EoS in \cite{Chatziioannou:2019yko} and a comparison of
the inferred radius in \cite{Chen:2019rja} together with a collection of
simulated BNS merger events for purely hadronic and hybrid EoS, it has
been shown that $\sim$ 50-100 detections can be sufficient to
distinguish different EoSs. In \cite{Coupechoux:2023fqq}, the authors assess the possibility to distinguish different equation of state models with and without a PT from a GW170817-like event during the O4 run of LVK. 

The above studies, however, conclude on the detectability of a
PT by comparing a few EoS models, which is clearly not sufficient to
cover the entire space of all possible EoSs with and without PT. A
different approach has been used in \cite{Landry:2020vaw,Essick:2020flb},
where a non-parametric EoS inference is applied to the GW170817 event
with weak statistical evidence in favor of two stable branches,
i.e. the existence of hybrid stars with a strong PT. In the study by
\cite{Pang:2020ilf}, the authors have performed a Bayesian inference study
with three different injected EoS models, concluding that already 12
events with current detectors could be sufficient to disentangle a
strong PT. However, the injected models represent only
snapshots of all possible EoSs with a PT transition and the inference
of a PT is done based on the number of different polytropes employed
in the EoS reconstruction. Here, we will use a meta-modelling approach
as well for the models with PT as for those without. The latter is a
flexible parameterisation of the purely nuclear EoS, incorporating
constraints from nuclear experiments, theory and astrophysical
observations as priors~\cite{Margueron18a}. For the former, we have
extended the nuclear meta-model to include a potential PT taking as
parameters the density for the onset of the transition, the energy
density jump and the sound speed in the high-density phase controlling
the stiffness of the EoS. The injection models have been chosen from
the borders of the respective meta models. In addition, we will
concentrate here on the possibility to detect a PT from
one single loud event in third-generation detectors.

The paper is organised as follows: In section \ref{sec:metamodel} we
summarise the EoS modelling which is used in the present work. A brief
summary of the Fisher matrix formalism to quantify variances in
various astrophysical parameters connected to a binary neutron star
(BNS) merger is provided in section \ref{sec:gwbench}. 
Section \ref{sec:bayes} contains the proposed Bayesian framework to infer the possible signs of a PT from the gravitational wave signal generated by a BNS coalescence. We discuss the results in section \ref{sec:results}. Concluding remarks and the future extensions of the present work are discussed in section \ref{sec:summary}.


\section{Metamodelling of neutron star matter}
\label{sec:metamodel}

For our purposes, NS matter comprises an inhomogeneous crust and a core of uniform matter fully governed by strong interaction. The crust and the purely nucleonic outer core are consistently calculated using the same energy functional, see section~\ref{subsec:crust}. We describe then the core, see section~\ref{subsec:core} in the nucleonic hypothesis up to a certain density, beyond which we assume the appearance of quarks at the very center of the star. The density of the phase transition to the quark core is an input parameter in our model. 

\subsection{Neutron star core}
\label{subsec:core}

\subsubsection{Nucleonic part}
\label{subsubsec:metaeos}

For the description of the purely nucleonic outer core, we follow a metamodelling approach. Below we will briefly recall the main lines of the model, for details see \cite{Margueron18a}. Within our model, the uniform nucleonic matter is described by decomposing the energy density of infinite nuclear matter as
\begin{eqnarray}
    \label{eq:meta-def}  
    \varepsilon_N(n_n,n_p)=C_{kin}\sum_{q=n,p}\dfrac{n_q^{5/3}}{m^\star_q(n,\delta)} +U_0(n)+U_{sym}(n) \, \delta^2 \, , 
\end{eqnarray}
where $n_n$ $(n_p)$ is the neutron (proton) number density, $n=n_n+n_p$ is the baryon number density, $\delta=(n_n-n_p)/n$ is the isospin asymmetry and $C_{kin}=3(3\pi^2\hbar^3)^{2/3}/10 \approx 2.87 \, \hbar^2$. 
The first term takes into account the zero point Fermi gas contribution, and the momentum dependence of the nuclear interaction through the density-dependent effective mass $m_q^\star$ $(q=n,p)$. 
The remaining term is broken down into an isospin symmetric part, $U_0(n)$, and an isospin asymmetric part, $\delta^2 \, U_{sym}(n)$. 
For both $U_0(n)$ and  $U_{sym}(n)$ an expansion up to  order $N=4$ around the nuclear saturation density $n_{sat}$ is used:
\begin{eqnarray}
\label{eq:meta-pot}
U_{0,sym}(n)=n\sum_{k=0}^N \dfrac{(v_k)_{0,sym}}{k!} \, u^{N}_{k}(x) \, x^k \, ,
\end{eqnarray}
with
\begin{equation}
    x = \dfrac{n - n_{sat}}{3n_{sat}},
 \text{\ \   and,\ \   }
u^{N}_{k}(x) = 1 - (-3x)^{N+1-k} \, e^{-b(1+3x)},
\end{equation}
where the $u^{N}_{k}$ terms ensure vanishing energy at zero density.
The coefficients $(v_k)_{0,sym}$ are functions of nuclear matter properties (NMPs) at saturation density $n_{sat}$.  Specifically, 
the coefficients of $U_0$ in Eq. \eqref{eq:meta-pot} can be expressed as a function of the density derivatives at different orders of the energy per baryon of symmetric nuclear matter\footnote{
    Nuclear matter with an equal number of protons and neutrons, i.e. $\delta=0$.
    } 
(SNM), like the energy per particle $E_{sat}$, incompressibility $K_{sat}$, skewness $Q_{sat}$ and kurtosis $Z_{sat}$. 
Similarly, the coefficients of $U_{sym}$ are related to other NMPs, 
like the symmetry energy $E_{sym}$, symmetry slope $L_{sym}$, symmetry incompressibility $K_{sym}$, symmetry skewness $Q_{sym}$ and symmetry kurtosis $Z_{sym}$, 
all evaluated at $n_{sat}$ and for symmetric matter.

The contribution of an ideal gas of electrons ($e$) and muons ($\mu$) to the energy density
\begin{eqnarray}
\label{eq:elec-ener}
  \varepsilon_{e,\mu} = \dfrac{(m_{e,\mu})^4}{8\pi^2(\hbar c)^3}\left[(2x_r^2+1)x_r\gamma_r-ln(x_r+\gamma_r)\right],
\end{eqnarray}
where, $x_r = \dfrac{\hbar c (3\pi^2 n_l)^{1/3}}{m_l}$ and $\gamma_r = \sqrt{1+x_r^2}$ with $l=e,\mu$.
The net amount of electrons present in the system is determined by the $\beta$-equilibrium condition solving the coupled equations of chemical potentials $\mu_{n,p,e}$ of neutrons, protons and electrons, respectively, as
\begin{eqnarray}
    \mu_n(n,\delta)-\mu_p(n,\delta)=\mu_e(n,\delta),
    \label{eq:betaeq}
\end{eqnarray}
where $\mu_{n,p}={\partial\varepsilon_N}/{\partial n_{n,p}}+m_{n,p}c^2$, and $m_{n,p}$ are the bare neutron and proton masses.
If the chemical potential of electrons 
$\mu_{e}=\partial \varepsilon_{e} / \partial n_{e} $ 
exceeds the muon mass $m_\mu$, they appear spontaneously in neutron star matter. 
Their amount is fixed by local charge equilibrium $n_{\mu}=n_p-n_e$, and the global equilibrium condition $\mu_{\mu}=\mu_e$. 
Once the equilibrium composition $\delta(n)$ 
is determined at each density from Eq. \eqref{eq:betaeq}, the baryonic pressure is obtained from the thermodynamic relation
\begin{eqnarray}
    \label{beta_pressure}
     p_N(n)=\sum_{q=n,p} n_q \left ( \mu_q - m_q c^2 \right )- \varepsilon_N(n,\delta).
\end{eqnarray}
where $n_q$, $\mu_q$ and $\delta$ are all functions of $n$, while $m_q$ are the two bare masses of neutrons and protons.

\subsubsection{Quark matter EoS}
\label{subsubsec:pt}

We assume that the appearance of quarks in the inner core gives rise to a first-order phase transition, namely  two continuous branches $p_N(\varepsilon_N)$, $p_Q(\varepsilon_Q)$ separated by a finite jump $\Delta\varepsilon$  in energy density  at a given value $p_t$ of the baryonic pressure. We note that $\varepsilon_t$ is the highest value of the energy density in the nucleonic (N) phase, and $\varepsilon_t+\Delta\varepsilon$ represents the lowest energy density value in the quark phase.   
As a consequence, a discontinuity in the sound speed  $c_s$ between the nucleonic   and quark    phases is also expected. 
When the core gets converted to quarks completely, we assume it fixes $c_s$ all the way up to the center of the neutron star from the point of transition. 
In this simple picture, the EoS of the quark core can be specified by (see e.g. \cite{Zdunik2013,Alford:2017qgh}):
\begin{eqnarray}
    &p_Q(\varepsilon_Q)=c_s^2(\varepsilon_Q-\varepsilon_{t}),&\\
    &\mu_Q(p_Q)=\mu_{t}\left[1+\dfrac{1+c_s^2}{c_s^2}\dfrac{p_Q}{\varepsilon_{t}}\right]^{{c_s^2}/({1+c_s^2})},&\\
    &n_Q(p_Q)=n_{t}\left[1+\dfrac{1+c_s^2}{c_s^2}\dfrac{p_Q}{\varepsilon_{t}}\right]^{{1}/({1+c_s^2})}=n_{t}\left(\dfrac{\mu_Q}{\mu_{t}}\right)^{1/c_s^2}.&
\end{eqnarray}
 Here, $\mu_t$ represents the baryonic chemical potential at the transition point ($\mu_B=\mu_n$ in the nucleonic phase), and $n_t$ is the baryonic density at the onset of the transition. 
Three parameters $n_t, c_s^2$ and $\Delta \varepsilon$ are used on top of the parameters for the nucleonic metamodelling to describe the neutron star core. 
In \cite{Zhang:2023wqj} a similar approach is used to model the neutron star EoS in order to obtain constraints on a potential hadron-quark transition from neutron star observables, but the nucleonic part in their case is limited to one single EoS, obtained by fixing the NMPs to a fiducial set. 

\subsection{Neutron star Crust}
\label{subsec:crust}

We model the neutron star crust by using the compressible liquid drop model (CLDM) approximation of \cite{Carreau19}, which allows us to extend the metamodel for uniform nucleonic matter in Sec. \ref{subsubsec:metaeos}. 
In the CLDM, the bulk energy of a spherical nucleus with $A$ nucleons, $Z$ protons, radius $r_N$ and bulk density $n_i$ is described by using Eq. \eqref{eq:meta-def} as $E_{bulk}=A \varepsilon_N(n_i,1-2Z/A)/n_i$. 
The total binding energy of a finite nucleus is obtained by adding surface, curvature and Coulomb terms. 
The surface and curvature contributions are expressed in terms of surface and curvature tensions $\sigma_s$  and $\sigma_c$ as \cite{Ravenhall:1983bdb,Maruyama:2005vb,Newton:2011dw}
\begin{eqnarray}
E_{{surf}} + E_{{curv}} &=&4\pi r_N^2\left ( \sigma_s(Z/A) +
        \dfrac{2\sigma_c(Z/A)}{r_N}\right ), \\ \label{eq:interface}
        \text{with,\ \ \ \ \ \ }\sigma_s(x)&=&\sigma_0\dfrac{2^{4}+b_s}{x^{-3}+b_s+(1-x)^{-3}}, \\ \label{eq:surface}
\sigma_c(x)&=&5.5 \sigma_s(x) \dfrac{\sigma_{0,c}}{\sigma_0}(\beta-x).\label{eq:curvature}   
\end{eqnarray}

 Finally, the Coulomb term in a spherical Wigner-Seitz (WS) cell reads: 
\begin{eqnarray}
    {E}_{coul}   =  \dfrac{8}{3}\left (\pi e{Z\over A} n_i\right )^2r_N^5 \, \eta_{coul}
    \left(\dfrac{r_N}{r_{WS}}\right),
    \\ \label{eq:Ecoul}
    \text{with\ \ \ \ \ }\eta_{coul}(x)= \dfrac{1}{5}\left [ x^3+ 2 \left ( 1- \dfrac{3}{2}x \right ) 
        \right ]. 
        \label{eq:eta_Coul}
\end{eqnarray}
Here, $e$ is the elementary charge, $r_{WS}$ the radius of a WS cell, and $\eta_{coul}$ the function taking into account electron screening in the Coulomb energy. 
The parameters $\sigma_0$, $\sigma_{0,c}$, $b_s$ and $\beta$ are obtained by optimizing the agreement of nuclear masses in vacuum with the AME2016 mass table \cite{Wang17,DinhThi21a,DinhThi21b}. 
The EoS is eventually obtained by minimizing the energy of the WS cell by varying $A,Z,n_i, r_N,r_{WS}$, as well as the neutron gas density $n_g$ \cite{Carreau19,DinhThi21a,DinhThi21b}.

\section{Estimation of detectors' response: Fisher information formalism}
\label{sec:gwbench}

In order to assess the uncertainty in the parameter estimation associated with future observations of GWs from the coalescence of BNSs, we make use of the publicly available tool \textsc{gwbench}~\cite{Borhanian21}, which implements the Fisher information paradigm~\cite{Cutler:1994ys, Poisson95}. 
The main drawback of this approach is that it is valid for events with a high signal-to-noise ratio (SNR)~\cite{Valisnieri08}. 
On the other hand, its main advantage is the considerable increase in computational speed, as compared to standard Bayesian analysis. Here we give a summary of our procedure to compute a sequence of simulated GW events from BNSs and estimate the parameters that could be inferred from such events by the third-generation Earth-based GW detectors. 
For details about \textsc{gwbench}, the interested reader can refer to \cite{Borhanian21}. 

We model events of BNS coalescence with masses $m_1$ and $m_2$ by using the waveform templates implemented in \textsc{gwbench}, namely the ``TaylorF2 + tidal'' models~\cite{Wade14}. 
They depend on the following parameters: $\left( \mathcal{M}_c, \eta, \vec{\chi}_1, \vec{\chi}_2, \mathcal{D}_L, \iota, \tilde{\Lambda}, \delta \tilde{\Lambda} \right)$.
Here, $\eta = m_1 m_2/ ( m_1 + m_2 )^2 $, $\vec{\chi}_{1,2}$ are the dimensionless spin vectors of the two NSs, $\mathcal{D}_L$ is the luminosity distance, $\iota$ is the inclination angle of its orbital plane with respect to the line of sight. Finally, $\tilde{\Lambda}$ and $\delta \tilde{\Lambda}$ are two parameters containing the tidal deformabilities of both stars $\left( \Lambda_1, \Lambda_2 \right)$ and are defined as:
\begin{eqnarray}
  \tilde{\Lambda} &=  \frac{8}{13} \left[ \left( 1 + 7\eta - 31\eta^2\right)\left( \Lambda_1 + \Lambda_2 \right) + \right. \nonumber
  \\
   &+ \left.  \sqrt{1 - 4\eta}\, \left(1 + 9\eta - 11\eta^2 \right) \left( \Lambda_1 - \Lambda_2 \right) \right],  \label{e:def_lambda} 
   \\
  \delta \tilde{\Lambda} &=  \frac{1}{2} \left[ \sqrt{1 - 4\eta}\, \left( 1 - \frac{13272}{1319}\eta + \frac{8944}{1319} \eta^2 \right) \left( \Lambda_1 + \Lambda_2 \right)+ \right. \nonumber 
  \\
  &+  \left. \left( 1 - \frac{15910}{1319}\eta +\frac{32850}{1319}\eta^2 + \frac{3380}{1319}\eta^3 \right) \left( \Lambda_1 - \Lambda_2 \right) \right] . \label{e:def_dlambda}
\end{eqnarray}
We use different EoS models as described in Sec.~\ref{sec:metamodel} to relate each NS mass $\left( m_1, m_2 \right)$ to  the corresponding  tidal deformability parameter 
$\left( \Lambda_1, \Lambda_2 \right)$. 
The $M$--$\Lambda$ relation is obtained for each EoS by solving the Tolman-Oppenheimer-Volkov (TOV) equations, together with the differential equations for perturbed relativistic stars described by \cite{Hinderer08, Hinderer09}.
In the case where a phase transition appears in the NS, additional terms due to jumps in thermodynamic quantities must be taken into account. To do this, we follow the approach by \cite{Pereira20}. 

For each EoS, we assume fixed values for the spins and inclination ${\chi}_{1}=0.01$,  ${\chi}_{2}=0.005$, $\iota=45^\circ$ and inject a series of events into \textsc{gwbench}
, with chosen values for the chirp mass $\mathcal{M}_c$, the mass ratio $q = m_2 / m_1$ and the luminosity distance $\mathcal{D}_L$. $\tilde{\Lambda}$ and $\delta \tilde{\Lambda}$ are deduced from the $M$--$\Lambda$ relation for the specific EoS considered and Eqs. \eqref{e:def_lambda}, \eqref{e:def_dlambda}. 
The detector features are those of the projected third-generation ground-based ones: triangle-shaped Einstein Telescope \cite[ET, see][]{Punturo:2010zz} and two Cosmic Explorer detectors \cite[CE, see][]{Reitze:2019iox}. Details about their power spectral densities (PSD) and exact projected locations are given in Appendix~C of \cite{Borhanian21}. 
\textsc{gwbench} then returns estimates of measurement errors in the parameters of our waveform models. 
These estimates shall be used in the next sections, in particular for the case of $\Tilde{\Lambda}$. 
Concerning $\delta \tilde{\Lambda}$, this parameter only enters at the sixth post-Newtonian (6 PN) order in the waveform, meaning, it is subdominant with respect to the leading-order 5 PN tidal correction \cite{Wade14}. For this reason, it is not possible to obtain error-bound estimates of $\delta \tilde{\Lambda}$ within our approach; it can be however estimated using EoS-independent relations \cite{Chatziioannou18}.

\section{Bayesian framework}
\label{sec:bayes}

We use a Bayesian framework to quantify the compatibility of a simulated observation between a purely nucleonic and a hybrid (nucleons+quarks) neutron star core. For the different EoS, we employed the techniques outlined in Secs. \ref{subsubsec:metaeos} and \ref{subsec:crust} based on \cite{Dinh-Thi21, Mondal23}. 
To build the prior for the nucleonic metamodel, 12 NMPs corresponding to uniform matter
were varied randomly with a constant probability distribution over a wide domain.
Since the expansion in Eq.~\eqref{eq:meta-pot} is truncated at the fourth order, it is necessary to use different $Q_{(sat,sym)}$ and $Z_{(sat,sym)}$  below and above $n_{sat}$ to increase the reliability of the expansion over a large density range. In particular, it makes the EoS free of any fictitious correlations between observables connected more to the low and high-density regimes, respectively. Since these high-order NMPs have no contribution in Eq.~\eqref{eq:meta-def} at $n_{sat}$, this way of choosing different values for coefficients of the same order in the expansion does not induce any discontinuity in the energy, pressure or sound speed. This leads to a total of 16 parameters for the nucleonic EoS.

For the hybrid EoSs, containing nucleons and quarks, we obtained three families namely ``PT03'', ``PT04'' and ``PT05'' based on the density $n_t$ at the onset of the nucleon-quark transition fixed to 0.3, 0.4 and 0.5~fm$^{-3}$, respectively. For all hybrid models, the width of the plateau $\Delta\varepsilon$ is chosen by
imposing a random value for the lowest baryon density in the quark phase varying between $n_t$ and 1.5$n_t$. 
Finally, the squared sound speed $c_s^2$ is randomly varied between 0.1 and 0.9~$c^2$, where $c$ is the speed of light. 
These large variations were used to cover a large  space in the $p(\varepsilon)$ plane by our hybrid models. It should be kept in mind, that a constant sound speed description can of course not recover a more complicated behaviour inherent to  more sophisticated microscopic models for quark matter, see e.g. \cite{Kurkela:2009gj,Xu:2015jwa,Chen:2015mda,Zacchi:2015oma,Alvarez-Castillo19,Otto:2019zjy,Jokela:2020piw,Shahrbaf:2021cjz}, but the chosen range should be able to enclose most models such that we consider our assumptions as very conservative. 

It is quite important to mention here the particular motivation to keep $n_t$ fixed at distinct values, rather than varying it randomly within a range, too. Since a random $n_t$ would only introduce further uncertainty on top of the nucleonic EoS, the fully nucleonic metamodel would be systematically preferred irrespective of any observation. We thus target to identify the signature of nucleon-quark phase transition, based on its early or late appearance in terms of density.

\subsection{Obtaining an informed prior}\label{subsec:informed_prior}

On the whole, to perform the Bayesian analysis we have used 16 parameters $(N_p=16)$ for the nucleonic metamodel and 19 parameters $(N_p=19)$ for ``PT03'', ``PT04'' and ``PT05'' hybrid metamodels. 
The uninformed prior distribution $P_{prior}\mathbf{X})=\prod_{k=1}^{N_p}P_k({X_k})$ of the parameter set $\mathbf {X}\equiv \{X_k,k=1,\dots N_p\}$ is obtained with a flat uncorrelated distribution $P_k(X_k)$. To obtain a prior informed by different observations, the probability of each model is then conditioned by the likelihood models of the AME2016 mass evaluation \cite{Wang17}, low density constraints on SNM and PNM obtained from theoretical $\chi$-EFT calculations \cite{Drischler16}, constraint from the observed maximum mass of NS \cite{Demorest:2010bx, Antoniadis:2013pzd} and constraints on the joint tidal deformability $\Tilde{\Lambda}$ of the GW170817 event \cite{LIGOScientific:2018hze} as
\begin{eqnarray}\label{eq:prob-informed}
    &&P_{prior}^{\text{informed}}(\mathbf{X})=P(\mathbf{X}|\mathbf{c})=\mathcal{N}\cdot P_{AME2016}(\mathbf{X})P_{\chi-EFT}(\mathbf{X})\nonumber\\
    &&\ \ \ \ \ \ \ \ \ \ \ \ \ \times\  P_{M_{max}}(\mathbf{X})P_{GW170817}(\mathbf{X})
    \prod_k P(c_k|\mathbf{X}).
\end{eqnarray}
Here, $\mathcal{N}$ is a normalization constant. The AME2016 filter is obtained as
\begin{eqnarray}\label{eq:filter-ame}
    P_{AME2016}(\mathbf{X})\propto \omega_{AME} \ \ e^{-\chi^2_{AME}(\mathbf{X})/2},
\end{eqnarray}
where, $\omega_{AME}=0$ or 1 depending on the meaningful reproduction or not, of the whole AME2016 mass table \cite{Wang17}, respectively. The objective function for the AME2016 mass table is given by 
\begin{eqnarray}
    \chi^2_{AME}({\bf X})=\dfrac{1}{N}\sum_n \dfrac{\left ( (B/A)_{CLDM}^{(n)}({\bf X})-(B/A)_{AME}^{(n)}\right )^2}{\sigma_{BE}^2},
\end{eqnarray}
with $N=2408$ and the adopted error $\sigma_{BE}=0.04$ MeV. The probability for the $\chi$-EFT pass-band type filter based on the constraints on SNM and PNM between $n=0.02$ and 0.2 fm$^{-3}$ is obtained from the theoretical calculation in  \cite{Drischler16} as,
\begin{eqnarray}\label{eq:prob-chieft}
    P_{\chi\text{-}EFT}(\mathbf{X}) \propto \omega_{\chi\text{-}EFT}(\mathbf{X}), 
\end{eqnarray}
where, $\omega_{\chi\text{-}EFT}(\mathbf{X})=0$ or 1, depending on SNM and PNM corresponding to $\mathbf{X}$ passing through the whole range or not. This theoretical band was obtained from the 90\% confidence interval, which we have increased by 5\% on the edges to interpret it as a $2\sigma$ band. The probability assigned to each model due to observed maximum mass $M_{max}^{obs}=2.01\pm 0.04\ M_{\odot}$ \cite{Antoniadis:2013pzd}, following a cumulative probability distribution, is given by
\begin{eqnarray}\label{eq:prob-mmax}
    P_{M_{max}}(\mathbf{X})=\dfrac{1}{0.04\sqrt{2\pi}} \int_{0}^{M_{max}({\bf X})
/M_{\odot}}e^{-\dfrac{(x-2.01)^2}{2\times 0.04^2}}dx.
\end{eqnarray}
The effect of the joint tidal deformability $\Tilde{\Lambda}$ observed during the GW170817 event \cite{LIGOScientific:2018hze} on the different metamodels and hybrid metamodels are obtained from a two dimensional probability distribution $P_{GW170817}(\Tilde{\Lambda}(q),q)$ as,
\begin{eqnarray}\label{eq:prob-gw170817}
    P_{GW170817}(\mathbf{X})=\sum_i P_{GW170817}\left(\Tilde{\Lambda}(q^{(i)}),q^{(i)}\right).
\end{eqnarray}
Here, we have assumed a constant $\mathcal{M}_c=1.186\ M_{\odot}$ due to the small uncertainty in the observed chirp mass. For each model $\mathbf{X}$, we sampled $q\in[0.73,1]$ and interpolated the probabilities from the two dimensional distribution $P_{GW170817}(\Tilde{\Lambda}(q),q)$ to perform the sum in Eq. \eqref{eq:prob-gw170817}.

Once the informed prior probability for each model is obtained, the corresponding probability distribution for the observables is obtained by marginalizing over the range of parameters $\mathbf{X}\in[\mathbf{X}_{min},\mathbf{X}_{max}]$ as,
\begin{eqnarray}\label{eq:prob-obs}
    P^{\text{informed}}_{prior}(Y)=\prod_{k=1}^N \int_{X_k^{min}}^{X_k^{max}}dX_k P_{prior}^{\text{informed}}(\mathbf{X}) \delta\left (Y-Y({\mathbf X})\right ).
\end{eqnarray}

\subsection{Confronting the models with simulated ``observations''}\label{subsec:gwbench}

To identify the signature of a phase transition in the gravitational wave signal from a BNS coalescence, we examine the compatibility of the purely nucleonic metamodel and the hybrid metamodels subjected to a given event. As described in section \ref{sec:gwbench}, we consider a hypothetical BNS coalescing event specified by $\{\mathcal{M}_c^0,q_0,\tilde{\Lambda}_0\}$, where ${\mathcal {M}_c}$ is the chirp mass, $q=m_2/m_1$ the mass ratio, and $\Tilde{\Lambda}$ the tidal deformability. To determine $\Tilde{\Lambda}_0(\mathcal{M}_c^0,q_0)$ we use a specific EoS model from one of the hybrid metamodel families. Given the characteristics of the detector, the distance ($\mathcal{D}_L$) and the sky location, we then calculate the interferometer response to this event via \textsc{gwbench} using the aforementioned EoS model.  This  gives us a posterior experimental distribution $p_{GW}^0(\mathcal{M}_c,q,\Tilde\Lambda)$, as well as the marginalized distributions $p_{GW}^0(\mathcal{M}_c)$, $p_{GW}^0(q)$, $p_{GW}^0(\tilde{\Lambda})$, that of course will implicitly depend on the choice $\mathcal{M}_c^0,q_0$, together with the detector characteristics. Since the chirp mass is very well measured, we will always assume $p_{GW}^0(\mathcal{M}_c)= \delta(\mathcal{M}_c-\mathcal{M}_c^0)$.

Once an ``observation'' is simulated with a model from the hybrid metamodel class, we want to confront the tidal polarizability measurement $p_{GW}^0(\tilde{\Lambda})$ with the nucleonic hypothesis. To make the comparison, in principle, we could calculate
\begin{eqnarray}\label{eq:prob-meta0}
    p_{meta}^{(0)}(\tilde{\Lambda})\equiv p\left(\tilde{\Lambda}\left|\right.meta, BI=\mathcal{M}_c^0,q_0\right),
\end{eqnarray}
where $meta$ denotes the nucleonic metamodel, and $BI$ the background information. However, if we consider that $p_{GW}^0(\tilde{\Lambda})$ corresponds to a true measurement, $\mathcal{M}_c^0$ and $q_0$ are not known exactly, but only as a distribution $p_{GW}^0(\mathcal{M}_c,q)$. Thus, the only quantity we can meaningfully compare to $p_{GW}^0(\tilde{\Lambda})$
is the distribution given by,
\begin{eqnarray}\label{eq:prob-meta1}
    p_{meta}^{(1)}(\tilde{\Lambda})\equiv p\left(\tilde{\Lambda}\left|\right.meta, BI=\mathcal{M}_c^0,p_{GW}^0(q)\right).
\end{eqnarray}
Imposing to the nucleonic metamodel a $q$ distribution identical to the one hypothetically extracted from the ``observation'' clearly gives a distribution more spread than $p_{meta}^{(0)}(\tilde{\Lambda})$, which is the only one that we will be able to compare to the observation. A similar probability distribution corresponding to the hybrid metamodel is expressed as
\begin{eqnarray}\label{eq:prob-ptmeta1}
    p_{PT}^{(1)}(\tilde{\Lambda})\equiv p\left(\tilde{\Lambda}\left|\right.PT, BI=\mathcal{M}_c^0,p_{GW}^0(q)\right),
\end{eqnarray}
where $PT$ signifies hybrid metamodels containing a first-order hadron-quark phase transition. 

In the end, to distinguish the compatibility of observation with the nucleonic metamodel and the hybrid metamodels, one can resort to evidence in terms of Bayes factors. Given an event, the  Bayes factor can be defined as a function of $\Tilde{\Lambda}$ as
\begin{eqnarray}\label{eq:bayes-lambda}
    B_{PT, meta}(\tilde{\Lambda})=\dfrac {p_{PT}^{(1)}(\tilde{\Lambda})}{p_{meta}^{(1)}(\tilde{\Lambda})},\\
    B_{meta, PT}(\tilde{\Lambda})=\dfrac{1}{B_{PT, meta}(\tilde{\Lambda})}.
\end{eqnarray}
The average Bayes factors associated to the simulated observation $p_{GW}^0(\tilde{\Lambda})$ is specified as
\begin{eqnarray}\label{eq:avg-bayes}
    \log\left(\langle B\rangle_{PT,meta}^{\mathcal{M}_c^0,q_0}\right) ={\int d\tilde{\Lambda}\; p_{GW}^0(\tilde{\Lambda}) 
        \log\left[\dfrac {p_{PT}^{(1)}(\tilde{\Lambda})}{p_{meta}^{(1)}(\tilde{\Lambda})}\right]},\\
\log\left(\langle B\rangle_{meta,PT}^{\mathcal{M}_c^0,q_0}\right) =\int d\tilde{\Lambda}\; p_{GW}^0(\tilde{\Lambda}) 
        \log\left[\dfrac {p_{meta}^{(1)}(\tilde{\Lambda})}{p_{PT}^{(1)}(\tilde{\Lambda})}\right]. 
        \label{eq:avg-bayes_rev}
\end{eqnarray}

\section{Results}
\label{sec:results}

\subsection{Hybrid metamodels}
\label{subsec:results_eos}

\begin{figure}
    \centering
    \includegraphics[width=0.48\textwidth]{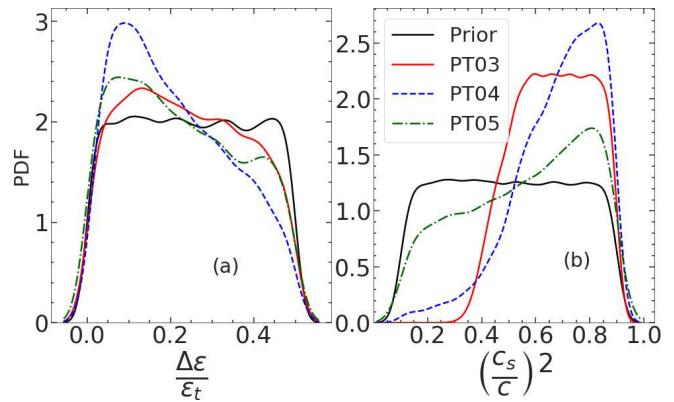}
    \caption{Informed prior distribution of the width of plateau $\Delta\varepsilon$ in units of energy density at transition $\varepsilon_t$ and squared sound speed $c_s^2$ in units of speed of light $c$ with phase transitions at 0.3, 0.4 and 0.5 fm$^{-3}$ for hybrid metamodels PT03, PT04 and PT05, respectively (see text for details).}
    \label{fig:pt_properties}
\end{figure}

To set the stage for confronting simulated observations with families of EoS with or without a PT to quark matter, we need to obtain 
an informed prior as described by Eq.\eqref{eq:prob-informed} and \eqref{eq:prob-obs}. Distributions of these informed priors for different nuclear matter parameters of the nucleonic metamodelling can be found in  \cite{Dinh-Thi21,Mondal23} as posteriors. In Fig. \ref{fig:pt_properties} we display the distributions of the sound speed parameter $c_s^2$ in panel (a) and width of the plateau $\Delta\varepsilon$ in panel (b) for the new class of hybrid metamodels PT03, PT04 and PT05, as described in Sec. \ref{subsubsec:pt} and \ref{sec:bayes}. To highlight the impact of the astrophysical constraints like the observed $M_{max}$ or $\tilde\Lambda$ from GW170817 on the hybrid metamodels, we display the prior distributions of $\Delta\varepsilon$ and $c_s^2$ in Fig. \ref{fig:pt_properties}, too. For PT03, models with small values of $c_s^2$ $(<0.4c^2)$ get suppressed significantly, but the rest still remain uniformly distributed. The hybrid metamodel PT04 clearly prefers to have larger $c_s^2$, which somewhat evens out for PT05. The distributions of $\Delta\varepsilon$ for different hybrid metamodels obtain crests at smaller values (\textit{i.e.} no first-order PT) compared to the uninformed flat prior. 

\begin{figure}
    \centering
    \includegraphics[width=0.48\textwidth]{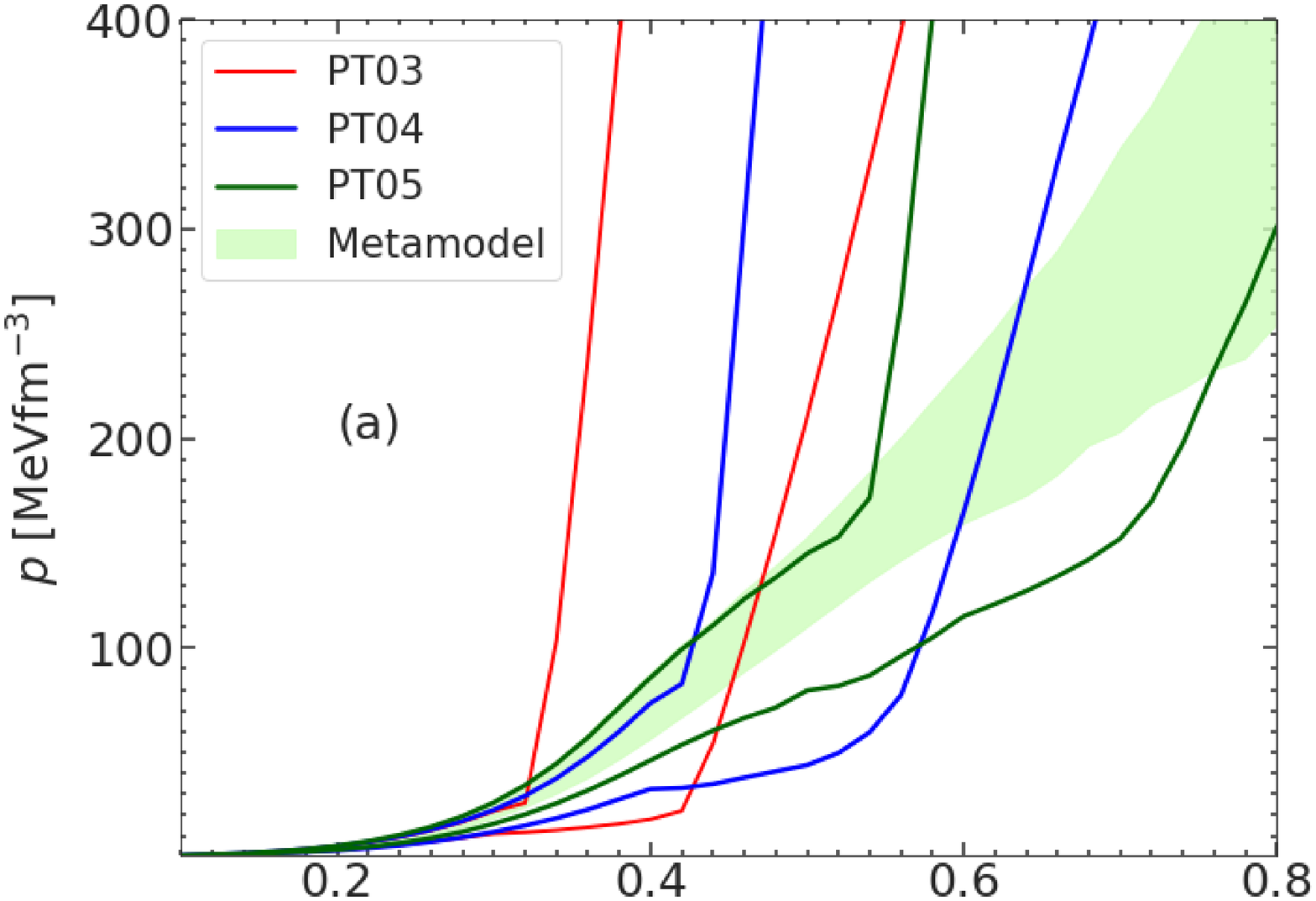}
    \includegraphics[width=0.48\textwidth]{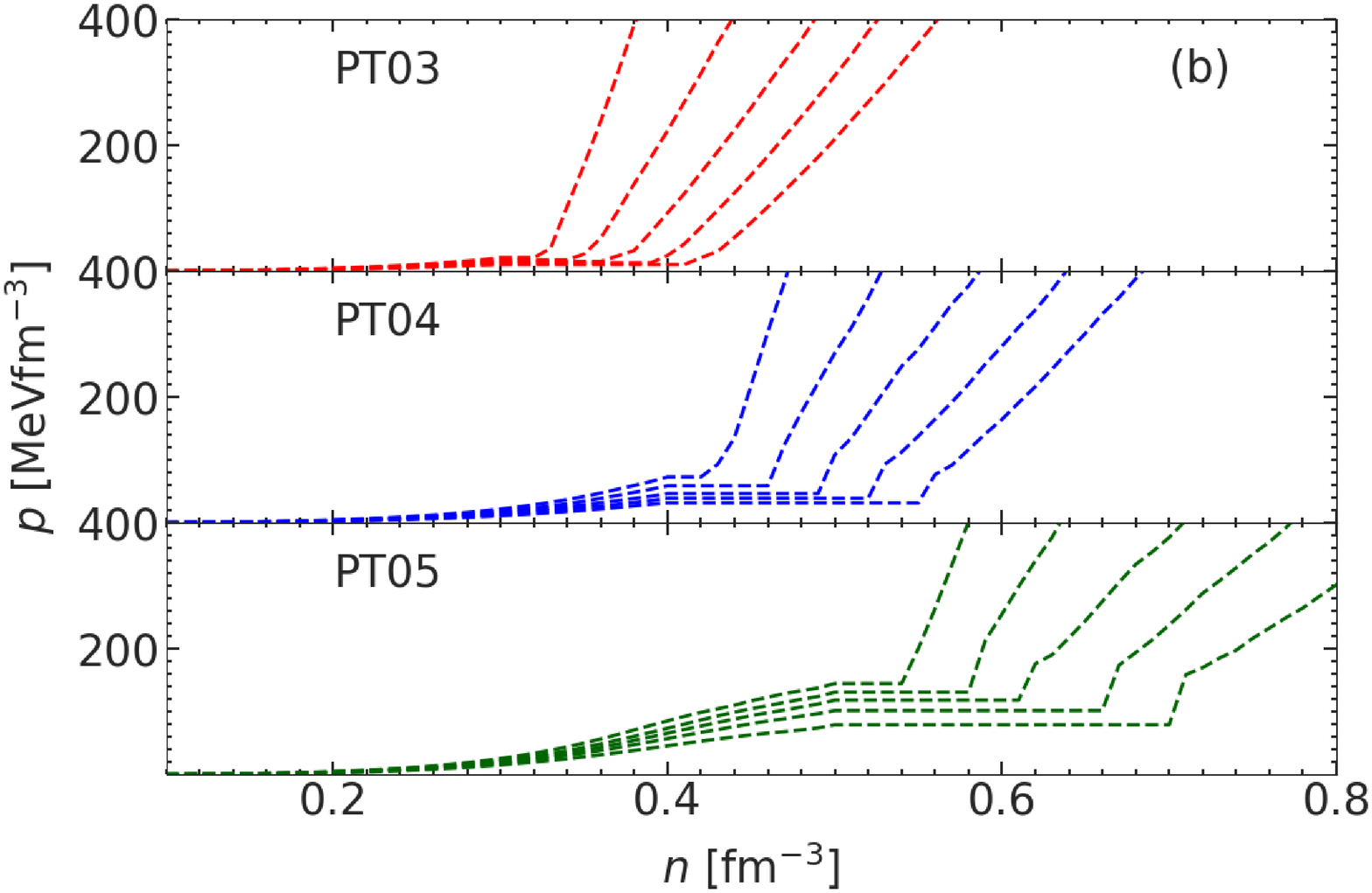}
    \caption{(a): Pressure at $\beta$-equilibrium as a function of number density $n$ at the 1$\sigma$ confidence interval obtained for nucleonic Metamodel and hybrid metamodels PT03, PT04 and PT05; (b): Injection models with first-order phase transitions based on the intervals given in panel (a) obtained for simulated ``observations'' using \textsc{gwbench} (see text for more details).}
    \label{fig:prho_pt}
\end{figure}

In Fig. \ref{fig:prho_pt}(a) we display the EoS of the nucleonic metamodel along with hybrid metamodels PT03, PT04 and PT05 in a 1$\sigma$ confidence interval. We  remind that these EoS posteriors are consistent with various observational constraints as described in section \ref{subsec:informed_prior}. One can observe that with increasing transition density $n_t$ from nucleonic to quark matter, the overall region explored in the $p$-$n$ plane increases. An early phase transition requires a very stiff behaviour of the quark phase, such as to meet the 2$M_{\odot}$ constraint. On the other hand, relatively softer behaviours, corresponding to lower values of the $c_s^2$ parameter, are allowed if the quark phase emerges at higher densities. This is consistent with the distribution of $c_s^2$ displayed in Fig. \ref{fig:pt_properties}(b). 

In order to simulate hypothetical observations within the Fisher formalism, one needs specific injection EoS models. We chose different injection models for PT03, PT04 and PT05 guided by the 1$\sigma$ boundaries displayed in Fig. \ref{fig:prho_pt}(a). 
The representative injection EoS are displayed in Fig. \ref{fig:prho_pt}(b) for PT03, PT04 and PT05 hybrid metamodels, respectively. Except for the plateau region, each individual injection model follows the 16\%, 33\%, 50\%, 67\% and 84\% quantiles of the corresponding posteriors from the bottom to the top (or right to left), respectively. We follow the nomenclature for the injection models obtained at the 16\% and 84\% quantiles as ``1$\sigma$-lower'' and ``1$\sigma$-upper'' models (c.f. Figs \ref{fig:ltilde_q}-\ref{fig:ltilde_distance}) of the corresponding hybrid metamodel classes \textit{i.e.} PT03, PT04 and PT05, respectively. 

\begin{figure}
    \centering
    \includegraphics[width=0.48\textwidth]{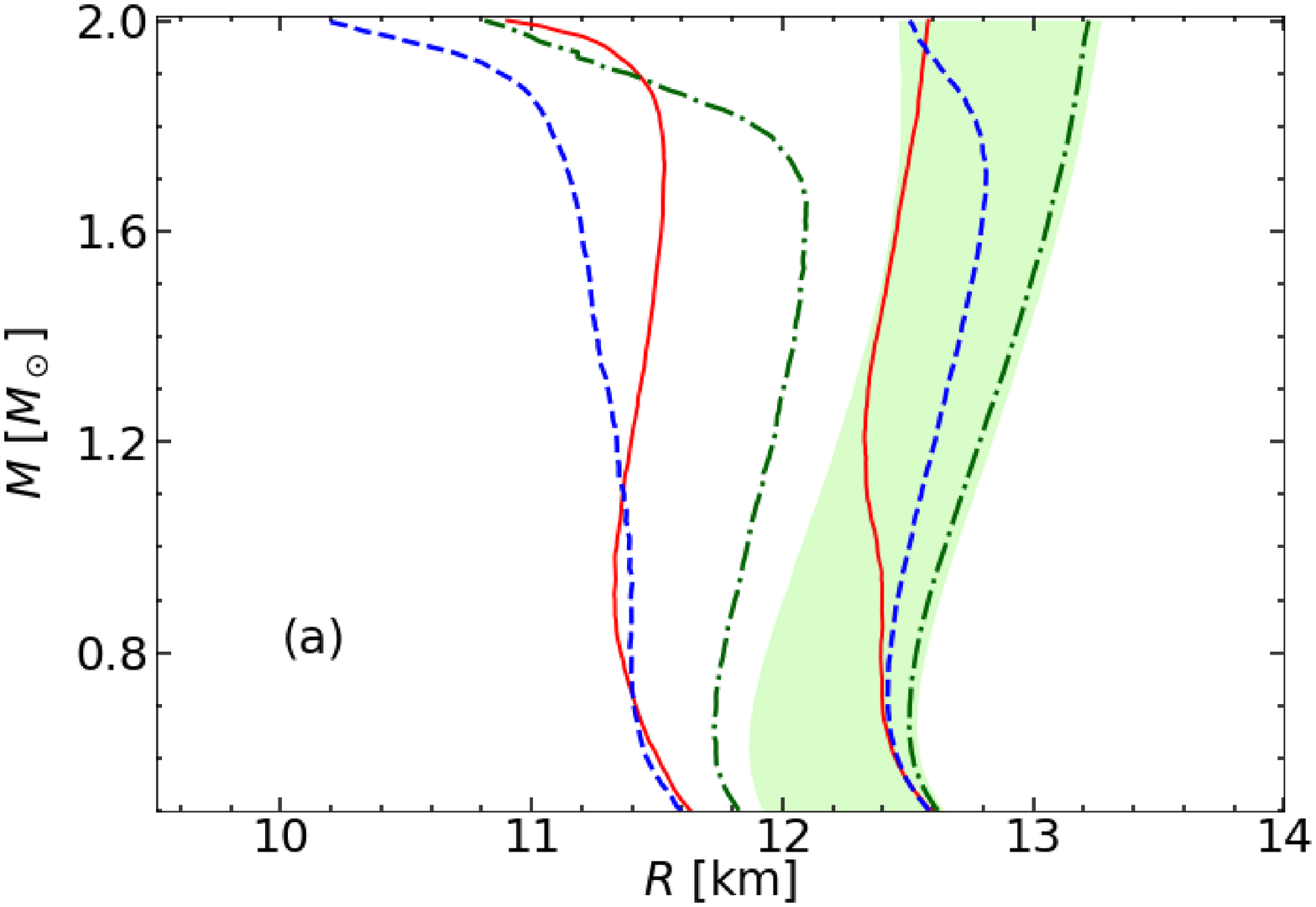}
    \includegraphics[width=0.48\textwidth]{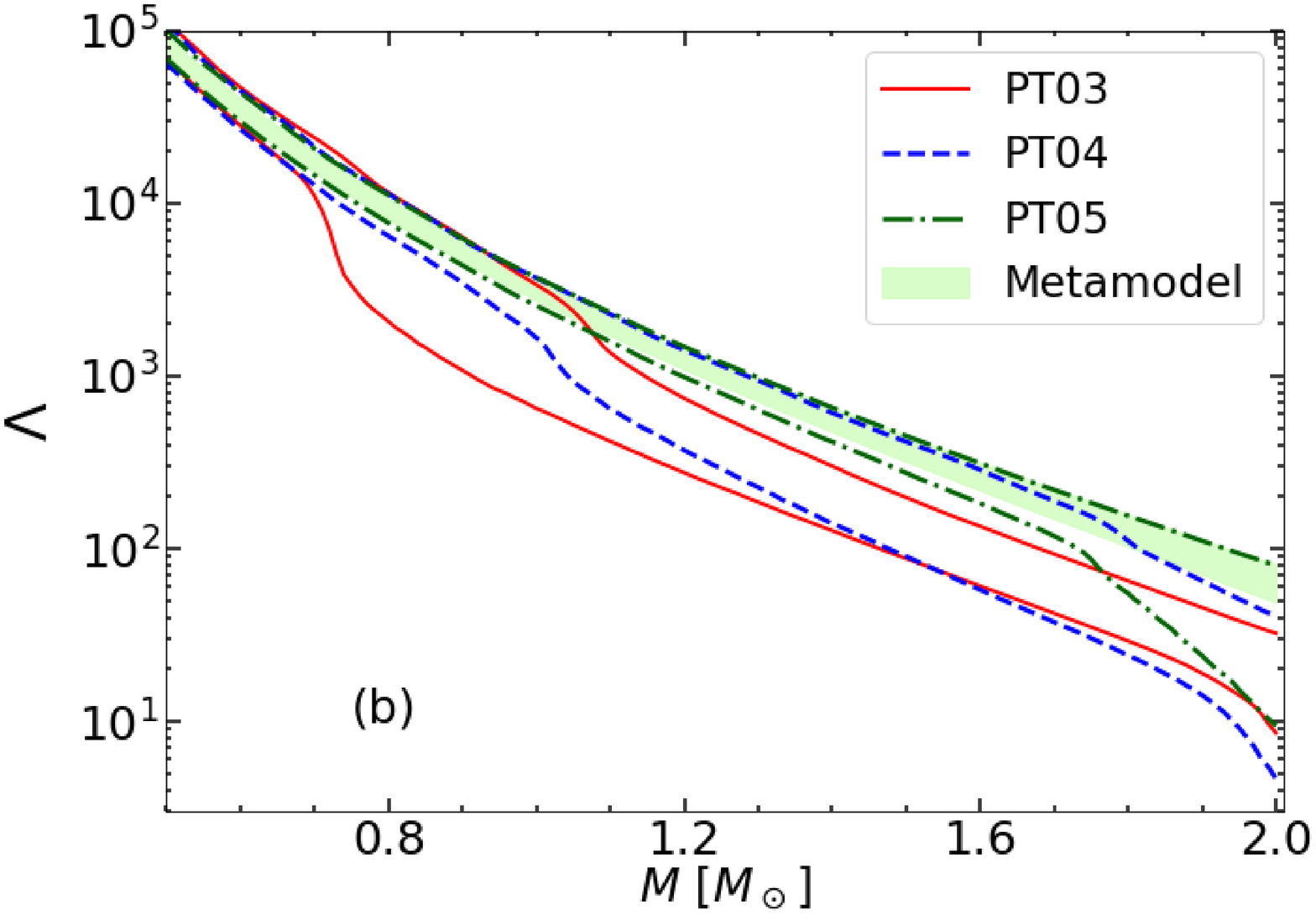}
    \caption{$M$-$R$ and $\Lambda$-$M$ relations at 1$\sigma$ confidence interval obtained for nucleonic Metamodel and hybrid metamodels PT03, PT04 and PT05.}
    \label{fig:mrl_pt}
\end{figure}

In Fig. \ref{fig:mrl_pt} we plot the two dimensional informed prior distribution for $M$-$R$ (panel(a)) and $\Lambda$-$M$ (panel (b)) relations of NSs using nucleonic and different hybrid metamodels. Introducing a first-order hadron-quark phase transition (c.f. Fig. \ref{fig:prho_pt}(a)) in the hybrid metamodels and the further requirements from various astrophysical constraints categorically soften the EoS. However, this does not systematically shift the hybrid metamodels to explore smaller $R$ or $\Lambda$ at a given $M$, with incremental $n_t$ \textit{i.e.} from 0.3 to 0.5 fm$^{-3}$ due to the behaviour of the nucleonic model in between the different transition densities. Beyond 1.2~$M_{\odot}$, PT03 and the nucleonic metamodel explore almost complementary regions in the $M$-$R$ and $\Lambda$-$M$ planes at the $1-\sigma$ level. The overlap regions increase significantly between PT04 and nucleonic Metamodel for both $M$-$R$ and $\Lambda$-$M$. At larger masses ($>1.5~M_{\odot}$) PT04 also explores smaller $R$ and $\Lambda$ compared to the other families of hybrid metamodels considered in this work. PT05 almost engulfs  the nucleonic metamodel, additionally exploring smaller values of $R$ and $\Lambda$ at larger masses ($>1.7\ M_{\odot}$). Essentially these differences in $\Lambda$ displayed in Fig. \ref{fig:mrl_pt} are manifested in the joint tidal deformability $\Tilde{\Lambda}$, which is explored in detail in the next section. 

\subsection{Simulated observations}
\begin{table}
\begin{center}
\caption{\label{tab:gebench_inj_par}
	Different \textsc{gwbench} injection parameters which were varied are listed. For all of them, ${\chi}_1=0.01$, ${\chi}_2=0.005$, $\iota=45^\circ$ were kept fixed. $\tilde{\Lambda}$ and $\delta \tilde{\Lambda}$ were fixed by the underlying injection models depicted in Fig. \ref{fig:prho_pt}(b).}
\begin{tabular}{ccc}
	\hline
 	\hline
  Parameter & Range & Step-size\\
  \hline
	$\mathcal{M}_c^0$ ($M_{\odot}$) & 1.1 - 1.46  & 0.045 \\
   \hline
    $q_0=\frac{m_2}{m_1}$ & 0.79 - 0.99 & 0.05\\
      \hline
$\mathcal{D}_L$ (Mpc) & 22, 120, 221, 326, 433, & \\
     & 544, 657, 772, 891, 1012 &  \\
	\hline
	\hline
\end{tabular}
\end{center}
\end{table}
As an exploratory study, we start by considering single events, choosing at each time specific values of chirp mass $\mathcal{M}_c^0$, mass ratio $q_0$, and luminosity distance $\mathcal{D}_L$ keeping the remaining parameters defining a BNS coalescence, \textit{viz.} spin $\Vec{\chi}$ of the constituents, inclination $\iota$ etc. fixed. Altogether, we considered 9 values of $\mathcal{M}_c^0$, 5 values of $q_0$ situated at 10 different $\mathcal{D}_L$ as outlined in Table. \ref{tab:gebench_inj_par}. The response of the Fisher matrix formalism for a chosen EoS was tested for these 450 assumed events with equal probability. 5 different hybrid EoSs from each of the PT03, PT04 and PT05 families were used for this purpose, which are displayed explicitly in Fig. \ref{fig:prho_pt}(b). Note that we use the notation $\mathcal{M}_c$ and $q$ instead of $\mathcal{M}_c^0$ and $q_0$, from here onward, respectively. This is just for simplicity. 

We want to emphasize here that assumptions about the chosen events, and how they are distributed in the sky can affect the quantitative outcomes we are going to demonstrate in this section. However, the methodology proposed here can be used to incorporate models of neutron star mass distributions, see e.g. \cite{Ozel:2016oaf}. 
We leave it as a 
future study. 

\begin{figure}
    \centering
    \includegraphics[width=0.48\textwidth]{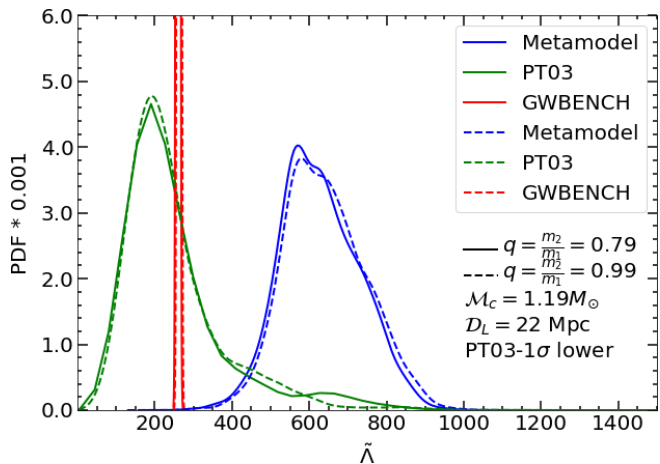}
    \caption{Probability distribution functions of joint tidal deformability $\tilde\Lambda$ in nuclonic Metamodel, hybrid metamodel PT03 and \textsc{gwbench} for mass ratio $q=0.79$ (solid) and $q=0.99$ (dashed) obtained using PT03-1$\sigma$-lower injection model, $\mathcal{M}_c=1.19\ M_{\odot}$ and $\mathcal{D}_L=22$ Mpc.}
    \label{fig:ltilde_q}
\end{figure}

Before diving into the calculation of Bayes factors as demonstrated in Eqs. \eqref{eq:avg-bayes}-\eqref{eq:avg-bayes_rev}, let us first analyze the premises chosen in the different astrophysical parameters and EoS modelling for the present endeavour. To this aim, we compare systematically the probability distribution functions (PDFs) of $\Tilde{\Lambda}$ of BNS merging events that enter in Eqs. \eqref{eq:avg-bayes}-\eqref{eq:avg-bayes_rev}, \textit{i.e.}, $p^0_{GW}(\Tilde{\Lambda})$,  $p^{(1)}_{PT}(\Tilde{\Lambda})$, and $p^{(1)}_{meta}(\Tilde{\Lambda})$ calculated from \textsc{gwbench}, hybrid and nucleonic metamodels, respectively, changing one variable at a time, keeping the rest fixed. In Fig. \ref{fig:ltilde_q}, we plot the PDFs of $\Tilde{\Lambda}$ calculated from nucleonic metamodel (blue), hybrid metamodel PT03 (green) and \textsc{gwbench} (red) for two extreme mass ratios $q=0.79$ (solid lines) and $q=0.99$ (dashed lines) considered in the present calculation. In both cases, $\mathcal{M}_c=1.19\ M_{\odot}$, $\mathcal{D}_L=22$ Mpc were kept fixed and the same injection model PT03-1$\sigma$-lower was used in \textsc{gwbench}. The differences due to the variation in $q$ are almost negligible.
In the case of a very close detection shown in Fig. \ref{fig:ltilde_q}, the theoretical uncertainties clearly prime over the observational ones. 
In spite of those large uncertainties, the two theoretical distributions only marginally overlap. The Bayes factors calculated for $q=0.79$ and $q=0.99$ as depicted in Fig. \ref{fig:ltilde_q} using Eq. (\ref{eq:avg-bayes}) turned out to be $\log \left(\langle B\rangle_{PT03,meta}^{1.19\ M_{\odot},0.79}\right)=2.15$ and $\log \left(\langle B\rangle_{PT03,meta}^{1.19\ M_{\odot},0.99}\right)=1.56$, respectively. 
We have observed that the quantitative values of the Bayes factors demonstrated here can further increase if a larger $\mathcal{M}_c$ is chosen.  


\begin{figure}
    \centering
    \includegraphics[width=0.48\textwidth]{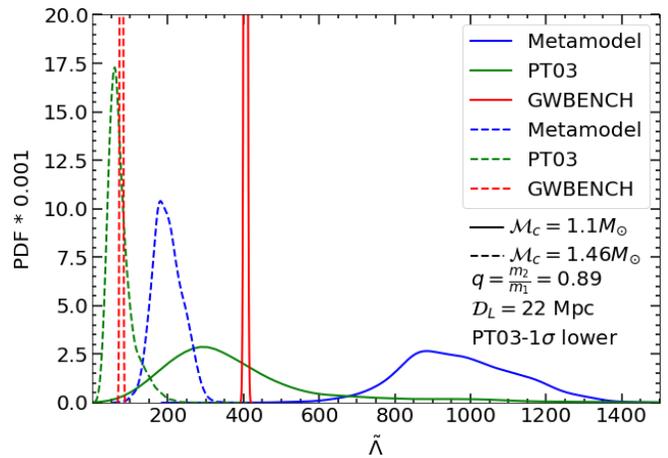}
    \caption{ The same as Fig. \ref{fig:ltilde_q} but for $\mathcal{M}_c=1.1\ M_{\odot}$ (solid) and $\mathcal{M}_c=1.46\ M_{\odot}$ (dashed) using PT03-1$\sigma$-lower injection model, $q=0.89$ and $\mathcal{D}_L=22$ Mpc.}
    \label{fig:ltilde_mc}
\end{figure}

This is exactly what is analyzed in Fig. \ref{fig:ltilde_mc}, where we show the PDFs of $\Tilde{\Lambda}$ calculated from the nucleonic metamodel (blue) and one of its hybrid counterparts PT03 (green) along with \textsc{gwbench} (red) for two extreme cases of chirp masses $\mathcal{M}_c=1.1\ M_{\odot}$ (solid lines) and $\mathcal{M}_c=1.46\ M_{\odot}$ (dashed lines) considered in the present calculation. 
The same injection model PT03-1$\sigma$-lower was used in this study as in Fig. \ref{fig:ltilde_q}. The mass ratio $q$ and luminosity distance $\mathcal{D}_L$ was kept fixed at 0.89 and 22 Mpc, respectively. 
Clearly, the difference in $\mathcal{M}_c$ results in a change in the position of the peaks of the PDFs as well as their widths. The particular scenario depicted in Fig. \ref{fig:ltilde_mc} for $\mathcal{M}_c=1.1\ M_{\odot}$ and $\mathcal{M}_c = 1.46\ M_{\odot}$ results in a big difference in the Bayes factors, $\log \left(\langle B\rangle_{PT03,meta}^{1.1\ M_{\odot},0.89}\right)=1.54$ and $\log \left(\langle B\rangle_{PT03,meta}^{1.46\ M_{\odot},0.89}\right)=3.22$, respectively. In general, the effect of $\mathcal{M}_c$ is found to be much stronger in the Bayes factor compared to the mass ratio $q$, if the rest of the variables are kept fixed. Identification of the first-order PT will thus be more probable from the future observations for events with larger $\mathcal{M}_c$'s. 

\begin{figure}
    \centering
    \includegraphics[width=0.48\textwidth]{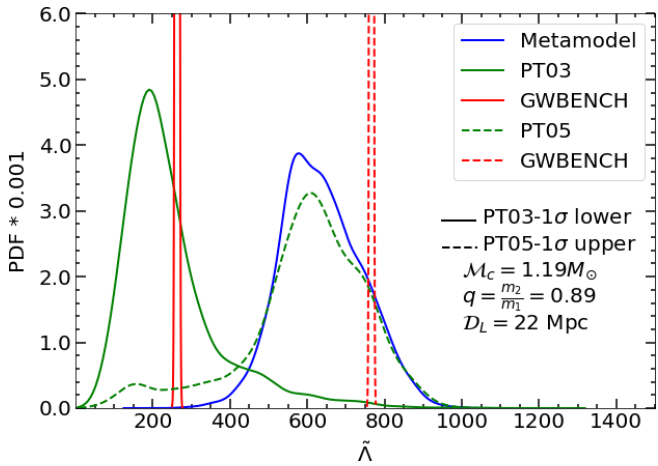}
    \caption{The same as Fig. \ref{fig:ltilde_q} but for injection models PT03-1$\sigma$-lower (solid) and PT05-1$\sigma$-upper (dashed) using  $\mathcal{M}_c=1.19\ M_{\odot}$, $q=0.89$ and $\mathcal{D}_L=22$ Mpc.}
    \label{fig:ltilde_inject}
\end{figure}

We lay our focus on the impact of injection models on the PDFs of $\Tilde{\Lambda}$ in Fig. \ref{fig:ltilde_inject}. Like the analysis done in Figs. \ref{fig:ltilde_q} and \ref{fig:ltilde_mc}, there is no exact guide to choose two extreme injection models from the ones outlined in Fig. \ref{fig:prho_pt}(b). Keeping an eye on Fig. \ref{fig:prho_pt}(a), we choose PT03-1$\sigma$-lower and PT05-1$\sigma$-upper with the idea that they have the least and maximum overlap with the nucleonic metamodel, respectively. PDFs of $\Tilde{\Lambda}$ obtained using PT03-1$\sigma$-lower (solid lines) and PT05-1$\sigma$-upper (dashed lines) are displayed in Fig. \ref{fig:ltilde_inject} for the nucleonic metamodel (blue), hybrid metamodels (green) and \textsc{gwbench} (red). One should note that, depending on the injection models we display the corresponding hybrid metamodels PT03 and PT05, respectively. Since the underlying astrophysical event is the same with $\mathcal{M}_c=1.19\ M_{\odot}$, $q=0.89$ and $\mathcal{D}_L=22$ Mpc, the PDF of $\Tilde{\Lambda}$ obtained with nucleonic Metamodel appears as a common one for both the injection models. The Bayes factors for the cases considered in Fig. \ref{fig:ltilde_inject} turn out to be $\log \left(\langle B\rangle_{PT03,meta}^{1.19\ M_{\odot},0.89}\right)=1.79$ and $\log \left(\langle B\rangle_{PT05,meta}^{1.19\ M_{\odot},0.89}\right)=0.02$ for PT03-1$\sigma$-lower and PT05-1$\sigma$-upper, respectively. This gives already an indication that identifying an early hadron-quark first-order PT will be more probable than a later one from future gravitational wave signals. 

\begin{figure}
    \centering
    \includegraphics[width=0.48\textwidth]{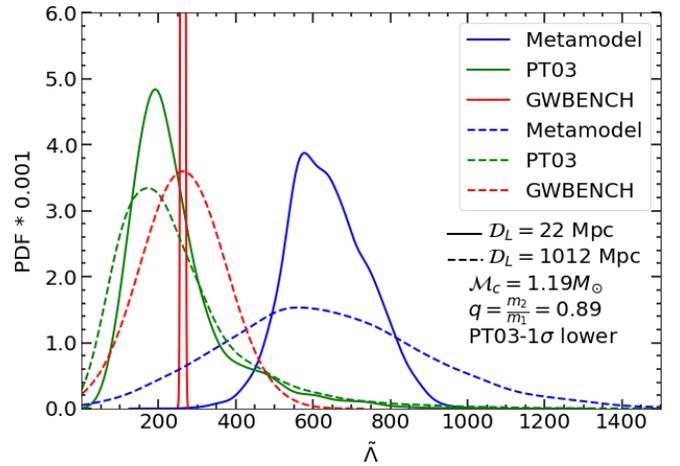}
    \caption{The same as Fig. \ref{fig:ltilde_q} but for $\mathcal{D}_L=22$ Mpc (solid) and $\mathcal{D}_L=1012$ Mpc (dashed) using PT03-1$\sigma$-lower injection model,  $\mathcal{M}_c=1.19\ M_{\odot}$ and $q=0.89$.}
    \label{fig:ltilde_distance}
\end{figure}

\begin{figure}
    \centering
    \includegraphics[width=0.48\textwidth]{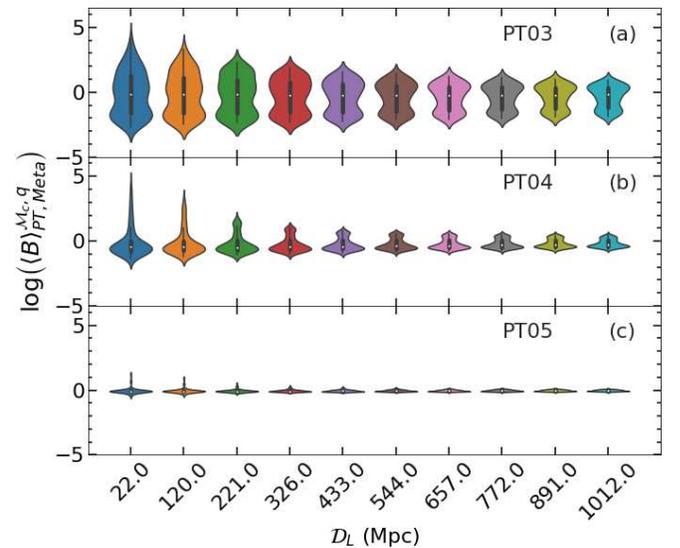}
    \caption{
    Average Bayes factor as a function of luminosity distance $\mathcal{D}_L$ for different injection models, averaged over the respective class of hybrid metamodels using Eq. \eqref{eq:avg-bayes}. The distributions come from the variation in $q$ and $\mathcal{M}_{c}$ considered in the present calculation (see Table \ref{tab:gebench_inj_par}).}
    \label{fig:avg_bayes}
\end{figure}

Depending on how far one BNS merger event takes place, the signal-to-noise ratio (SNR) recorded in gravitational wave interferometers can vary a lot. We have analyzed, to this aim, an event with $\mathcal{M}_c=1.19\ M_{\odot}$, $q=0.89$ simulated using PT03-1$\sigma$-lower injection model at luminosity distances $\mathcal{D}_L=22$ Mpc (solid lines) and $\mathcal{D}_L=1012$ Mpc (dashed lines) in Fig. \ref{fig:ltilde_distance}. The PDFs of $\Tilde{\Lambda}$ corresponding to the nucleonic metamodel, hybrid metamodel PT03 and \textsc{gwbench} are displayed in blue, green and red, respectively. The broadening of the PDFs from the case with $\mathcal{D}_L=22$ Mpc to the case with $\mathcal{D}_L=1012$ Mpc is clearly visible, which eventually affects the calculation of Bayes factors using Eq. \eqref{eq:avg-bayes}. For $\mathcal{D}_L=22$ Mpc, the Bayes factor is $\log \left(\langle B\rangle_{PT03,meta}^{1.19\ M_{\odot},0.89}\right)=1.79$ (same as the one obtained from the solid lines in Fig. \ref{fig:ltilde_inject}); for $\mathcal{D}_L=1012$ Mpc the Bayes factor becomes $\log \left(\langle B\rangle_{PT03,meta}^{1.19\ M_{\odot},0.89}\right)=0.76$. Even though in this particular case depicted in Fig. \ref{fig:ltilde_distance}, there is a clear hint of hybrid metamodel to be preferred over the nucleonic one, the evidence is not significant enough, even if we suppose an early phase transition. 

To have the global picture of the PT detectability, we plot the Bayes factors at different luminosity distances $\mathcal{D}_L$ comparing PT03, PT04 and PT05 against the nucleonic metamodel in panels (a), (b) and (c) of Fig. \ref{fig:avg_bayes}, respectively. In each panel, at a given $\mathcal{D}_L$, the variation in $\log \left(\langle B\rangle_{PT,meta}^{\mathcal{M}_c,q}\right)$ comes from the variation in $\mathcal{M}_c$, $q$ and injections models. Since we are interested in identifying the notion of a phase transition, high 
values of $\log \left(\langle B\rangle_{PT,meta}^{\mathcal{M}_c,q}\right)\ge 1-2$ 
is our primary concern \cite{Kass_bayesfactors_1995}. From Fig. \ref{fig:avg_bayes}(c) it is evident that the existence of a first-order transition at around  0.5 fm$^{-3}$ ($\sim 3n_{sat}$) 
can't be identified from a single gravitational wave signal with third generation interferometers. In Fig. \ref{fig:avg_bayes}(a) the other extreme case considered in the present calculation, \textit{i.e.}, a phase transition at 0.3 fm$^{-3}$ ($\sim 2n_{sat}$) seems to be a viable situation that can be identified particularly with high confidence at distances less than 300 Mpc. The highest positive values of $\log \left(\langle B\rangle_{PT,meta}^{\mathcal{M}_c,q}\right)$ at almost all distances are associated with larger $\mathcal{M}_c$'s obtained using injection models with lower quantiles (see Fig. \ref{fig:prho_pt}(b) and the corresponding discussion). It seems from Fig. \ref{fig:avg_bayes}(b) that only for a small percentage of events at low luminosity distances ($\mathcal{D}_L\lesssim 100$ Mpc) a phase transition can be identified if nature prefers to have it at $\sim 0.4$ fm$^{-3}$. Particularly, at $\mathcal{D}_L=22$ Mpc the high $\log \left(\langle B\rangle_{PT,meta}^{\mathcal{M}_c,q}\right)$ values correspond to $\mathcal{M}_c>1.4\ M_{\odot}$ obtained with PT04-1$\sigma$-lower injection model. 
We expect that the detectability of a first order phase transition reported in the present study will be largely improved by considering multiple detections, as expected by the future Einstein Telescope. To this aim, we plan to study the evolution of the Bayes factors as a function of the number of detections, by including realistic population distributions.


\section{Summary and discussion}
\label{sec:summary}

In summary, we have presented an updated metamodelling technique for the EoS in neutron star matter including potential first-order hadron-quark phase transitions. The hadronic core is assumed to have only nucleons and leptons up to a specified density. The EoS for the nucleonic core is obtained by an optimized expansion in number density truncated at  fourth order following \cite{Margueron18a}. 
The crust EoS and composition are subsequently extracted with a unified approach 
in the spherical Wigner-Seitz approximation. The extension of the EoS modelling in the quark core is done with the constant sound speed model, without dwelling on the microscopic composition. 

Three classes of hybrid (nucleon + quark) metamodels PT03, PT04 and PT05 named after the density corresponding to the hadron-quark phase transition, \textit{i.e.}, 0.3, 0.4 and 0.5 fm$^{-3}$, respectively, were generated. These hybrid metamodels were already made compliant with different nuclear physics and astrophysics constraints within the Bayesian paradigm. Using these hybrid and nucleonic posteriors,
we proposed a framework based on Bayes factors to discriminate a possible sign of phase transition from future gravitational wave signals generated by BNS mergers.  
To simulate future observational signals, we used the Fisher matrix formalism employing the publicly available tool \textsc{gwbench}~\cite{Borhanian21} that simulates the gravitational wave signal  using the TaylorF2 + tidal waveform templates and includes  the detector features of the projected third-
generation ground-based interferometers. 
We considered a single detection, and compared different chosen cases corresponding to different masses of NSs, located at different distances, and using injection models which include first-order phase transition. In particular, these hybrid injection models were constructed out of PT03, PT04 and PT05 1$\sigma$ posteriors, which are already informed by different physical constraints. We have critically assessed the impact of diversified variables in the discrimination of a phase transition signature through the Bayes factors. 

We have found that the mass ratio of the constituents of a BNS merger does not play a significant role in the magnitude of the Bayes factors. Overall, higher chirp mass, smaller luminosity distances and early phase transition with strong first-order effect can facilitate a possible identification of phase transition from future gravitational wave signals. Further, we have found that if nature prefers to have a phase transition at higher densities ($\gtrsim 3n_{sat}$), it is most likely to be masked, since it would be possible to explain that with an EoS model without phase transition. 

This study is directed mostly towards structuring a framework to look for the signatures of phase transition in future gravitational wave signals generated by BNS mergers. A further study incorporating realistic population models is needed to estimate the effect of multiple detections. 
It is also going to be important to incorporate microscopic modelling in the quark phase to extract further physical information regarding composition at high densities. We leave these ventures for future studies. 


\section{Acknowledgements}

We would like to thank Anthea Fantina for enlightening discussions as well as a careful reading of the manuscript, and Ssohrab Borhanian for fruitful exchanges and support on gwbench. 
This work has been partially supported by the IN2P3 Master Project NewMAC and the AAPG2022 ANR project GWsNS.
The authors gratefully acknowledge the Italian Instituto Nazionale de Fisica Nucleare (INFN), the French Centre National de la Recherche Scientifique (CNRS) and the Netherlands Organization for Scientific Research for the construction and operation of the Virgo detector and the creation and support of the EGO consortium.

\section{Data Availability}

No new data were generated in support of this research. The numerical results and the code underlying this article will be shared upon reasonable request to the authors.



\begin{thebibliography}{10}

\bibitem{haensel_book}
P.~Haensel, A.Y. Potekhin, and D.G. Yakovlev.
\newblock {\em Neutron Stars 1: Equation of State and Structure}.
\newblock Astrophysics and Space Science Library. Springer New York, 2007.

\bibitem{FiorellaBurgio:2018dga}
Fiorella~G. Burgio and Anthea~F. Fantina.
\newblock {Nuclear Equation of state for Compact Stars and Supernovae}.
\newblock {\em Astrophys. Space Sci. Libr.}, 457:255--335, 2018.

\bibitem{Oertel:2016bki}
M.~Oertel, M.~Hempel, T.~Kl\"ahn, and S.~Typel.
\newblock {Equations of state for supernovae and compact stars}.
\newblock {\em Rev. Mod. Phys.}, 89(1):015007, 2017.

\bibitem{Lattimer:2019iye}
J.~M. Lattimer.
\newblock {Impact of GW170817 for the nuclear physics of the EOS and the
  r-process}.
\newblock {\em Annals Phys.}, 411:167963, 2019.

\bibitem{Chamel:2008ca}
N.~Chamel and P.~Haensel.
\newblock {Physics of Neutron Star Crusts}.
\newblock {\em Living Rev. Rel.}, 11:10, 2008.

\bibitem{Raduta:2022elz}
Adriana~R. Raduta.
\newblock {Equations of state for hot neutron stars-II. The role of exotic
  particle degrees of freedom}.
\newblock {\em Eur. Phys. J. A}, 58(6):115, 2022.

\bibitem{Demorest:2010bx}
Paul Demorest, Tim Pennucci, Scott Ransom, Mallory Roberts, and Jason Hessels.
\newblock {Shapiro Delay Measurement of A Two Solar Mass Neutron Star}.
\newblock {\em Nature}, 467:1081--1083, 2010.

\bibitem{Antoniadis:2013pzd}
John Antoniadis et~al.
\newblock {A Massive Pulsar in a Compact Relativistic Binary}.
\newblock {\em Science}, 340:6131, 2013.

\bibitem{Fonseca:2016tux}
Emmanuel Fonseca et~al.
\newblock {The NANOGrav Nine-year Data Set: Mass and Geometric Measurements of
  Binary Millisecond Pulsars}.
\newblock {\em Astrophys. J.}, 832(2):167, 2016.

\bibitem{Cromartie:2019kug}
H.~Thankful Cromartie et~al.
\newblock {Relativistic Shapiro delay measurements of an extremely massive
  millisecond pulsar}.
\newblock {\em Nature Astronomy}, page 439, Sep 2019.

\bibitem{Miller:2021qha}
M.~C. Miller et~al.
\newblock {The Radius of PSR J0740+6620 from NICER and XMM-Newton Data}.
\newblock {\em Astrophys. J. Lett.}, 918(2):L28, 2021.

\bibitem{Riley:2021pdl}
Thomas~E. Riley et~al.
\newblock {A NICER View of the Massive Pulsar PSR J0740+6620 Informed by Radio
  Timing and XMM-Newton Spectroscopy}.
\newblock {\em Astrophys. J. Lett.}, 918(2):L27, 2021.

\bibitem{Miller:2019cac}
M.~C. Miller et~al.
\newblock {PSR J0030+0451 Mass and Radius from $NICER$ Data and Implications
  for the Properties of Neutron Star Matter}.
\newblock {\em Astrophys. J. Lett.}, 887(1):L24, 2019.

\bibitem{Riley:2019yda}
Thomas~E. Riley et~al.
\newblock {A $NICER$ View of PSR J0030+0451: Millisecond Pulsar Parameter
  Estimation}.
\newblock {\em Astrophys. J. Lett.}, 887(1):L21, 2019.

\bibitem{LIGOScientific:2017vwq}
B.~P. Abbott et~al.
\newblock {GW170817: Observation of Gravitational Waves from a Binary Neutron
  Star Inspiral}.
\newblock {\em Phys. Rev. Lett.}, 119(16):161101, 2017.

\bibitem{Punturo:2010zz}
M.~Punturo et~al.
\newblock {The Einstein Telescope: A third-generation gravitational wave
  observatory}.
\newblock {\em Class. Quant. Grav.}, 27:194002, 2010.

\bibitem{Maggiore:2019uih}
Michele Maggiore et~al.
\newblock {Science Case for the Einstein Telescope}.
\newblock {\em JCAP}, 03:050, 2020.

\bibitem{Reitze:2019iox}
David Reitze et~al.
\newblock {Cosmic Explorer: The U.S. Contribution to Gravitational-Wave
  Astronomy beyond LIGO}.
\newblock {\em Bull. Am. Astron. Soc.}, 51(7):035, 2019.

\bibitem{Evans:2021gyd}
Matthew {Evans}, Rana~X {Adhikari}, Chaitanya {Afle}, Stefan~W. {Ballmer},
  Sylvia {Biscoveanu}, Ssohrab {Borhanian}, Duncan~A. {Brown}, Yanbei {Chen},
  Robert {Eisenstein}, Alexandra {Gruson}, Anuradha {Gupta}, Evan~D. {Hall},
  Rachael {Huxford}, Brittany {Kamai}, Rahul {Kashyap}, Jeff~S. {Kissel}, Kevin
  {Kuns}, Philippe {Landry}, Amber {Lenon}, Geoffrey {Lovelace}, Lee
  {McCuller}, Ken K.~Y. {Ng}, Alexander~H. {Nitz}, Jocelyn {Read}, B.~S.
  {Sathyaprakash}, David~H. {Shoemaker}, Bram J.~J. {Slagmolen}, Joshua~R.
  {Smith}, Varun {Srivastava}, Ling {Sun}, Salvatore {Vitale}, and Rainer
  {Weiss}.
\newblock {A Horizon Study for Cosmic Explorer: Science, Observatories, and
  Community}.
\newblock {\em arXiv e-prints}, page arXiv:2109.09882, September 2021.

\bibitem{Glendenning:1991ic}
N.~K. Glendenning, F.~Weber, and S.~A. Moszkowski.
\newblock {Neutron and hybrid stars in the derivative coupling model}.
\newblock {\em Phys. Rev. C}, 45:844--855, 1992.

\bibitem{Alford:2001dt}
Mark~G. Alford.
\newblock {Color superconducting quark matter}.
\newblock {\em Ann. Rev. Nucl. Part. Sci.}, 51:131--160, 2001.

\bibitem{Buballa:2014jta}
Michael Buballa et~al.
\newblock {EMMI rapid reaction task force meeting on quark matter in compact
  stars}.
\newblock {\em J. Phys. G}, 41(12):123001, 2014.

\bibitem{Blaschke_new}
G.~A. Contrera, D.~Blaschke, J.~P. Carlomagno, A.~G. Grunfeld, and S.~Liebing.
\newblock Quark-nuclear hybrid equation of state for neutron stars under modern
  observational constraints.
\newblock {\em PHYSICAL REVIEW C}, 105(4), APR 27 2022.

\bibitem{Bauswein:2018bma}
Andreas Bauswein, Niels-Uwe~F. Bastian, David~B. Blaschke, Katerina
  Chatziioannou, James~A. Clark, Tobias Fischer, and Micaela Oertel.
\newblock {Identifying a first-order phase transition in neutron star mergers
  through gravitational waves}.
\newblock {\em Phys. Rev. Lett.}, 122(6):061102, 2019.

\bibitem{Most:2018eaw}
Elias~R. Most, L.~Jens Papenfort, Veronica Dexheimer, Matthias Hanauske, Stefan
  Schramm, Horst St\"ocker, and Luciano Rezzolla.
\newblock {Signatures of quark-hadron phase transitions in general-relativistic
  neutron-star mergers}.
\newblock {\em Phys. Rev. Lett.}, 122(6):061101, 2019.

\bibitem{Blacker:2020nlq}
Sebastian Blacker, Niels-Uwe~F. Bastian, Andreas Bauswein, David~B. Blaschke,
  Tobias Fischer, Micaela Oertel, Theodoros Soultanis, and Stefan Typel.
\newblock {Constraining the onset density of the hadron-quark phase transition
  with gravitational-wave observations}.
\newblock {\em Phys. Rev. D}, 102(12):123023, 2020.

\bibitem{Weih:2019xvw}
Lukas~R. Weih, Matthias Hanauske, and Luciano Rezzolla.
\newblock {Postmerger Gravitational-Wave Signatures of Phase Transitions in
  Binary Mergers}.
\newblock {\em Phys. Rev. Lett.}, 124(17):171103, 2020.

\bibitem{Torres-Rivas:2018svp}
Andoni Torres-Rivas, Katerina Chatziioannou, Andreas Bauswein, and
  James~Alexander Clark.
\newblock {Observing the post-merger signal of GW170817-like events with
  improved gravitational-wave detectors}.
\newblock {\em Phys. Rev. D}, 99(4):044014, 2019.

\bibitem{Branchesi:2023mws}
Marica Branchesi et~al.
\newblock {Science with the Einstein Telescope: a comparison of different
  designs}.
\newblock {\em arXiv e-prints}, page arXiv:2303.15923, 3 2023.

\bibitem{Damour:2009vw}
Thibault Damour and Alessandro Nagar.
\newblock {Relativistic tidal properties of neutron stars}.
\newblock {\em Phys. Rev. D}, 80:084035, 2009.

\bibitem{Postnikov:2010yn}
Sergey Postnikov, Madappa Prakash, and James~M. Lattimer.
\newblock {Tidal Love Numbers of Neutron and Self-Bound Quark Stars}.
\newblock {\em Phys. Rev. D}, 82:024016, 2010.

\bibitem{Sieniawska:2018zzj}
M.~Sieniawska, W.~Turczanski, M.~Bejger, and J.~L. Zdunik.
\newblock {Tidal deformability and other global parameters of compact stars
  with strong phase transitions}.
\newblock {\em Astron. Astrophys.}, 622:A174, 2019.

\bibitem{Han:2018mtj}
Sophia Han and Andrew~W. Steiner.
\newblock {Tidal deformability with sharp phase transitions in (binary) neutron
  stars}.
\newblock {\em Phys. Rev. D}, 99(8):083014, 2019.

\bibitem{Chatziioannou:2019yko}
Katerina Chatziioannou and Sophia Han.
\newblock {Studying strong phase transitions in neutron stars with
  gravitational waves}.
\newblock {\em Phys. Rev. D}, 101(4):044019, 2020.

\bibitem{Chen:2019rja}
Hsin-Yu Chen, Paul~M. Chesler, and Abraham Loeb.
\newblock {Searching for exotic cores with binary neutron star inspirals}.
\newblock {\em Astrophys. J. Lett.}, 893(1):L4, 2020.

\bibitem{Coupechoux:2023fqq}
J.~F. {Coupechoux}, R.~{Chierici}, H.~{Hansen}, J.~{Margueron},
  R.~{Somasundaram}, and V.~{Sordini}.
\newblock {Impact of O4 future detection on the determination of the dense
  matter equations of state}.
\newblock {\em arXiv e-prints}, page arXiv:2302.04147, February 2023.

\bibitem{Landry:2020vaw}
Philippe Landry, Reed Essick, and Katerina Chatziioannou.
\newblock {Nonparametric constraints on neutron star matter with existing and
  upcoming gravitational wave and pulsar observations}.
\newblock {\em Phys. Rev. D}, 101(12):123007, 2020.

\bibitem{Essick:2020flb}
Reed Essick, Ingo Tews, Philippe Landry, Sanjay Reddy, and Daniel~E. Holz.
\newblock {Direct Astrophysical Tests of Chiral Effective Field Theory at
  Supranuclear Densities}.
\newblock {\em Phys. Rev. C}, 102(5):055803, 2020.

\bibitem{Pang:2020ilf}
Peter T.~H. Pang, Tim Dietrich, Ingo Tews, and Chris Van Den~Broeck.
\newblock {Parameter estimation for strong phase transitions in supranuclear
  matter using gravitational-wave astronomy}.
\newblock {\em Phys. Rev. Res.}, 2(3):033514, 2020.

\bibitem{Margueron18a}
J\'er\^ome Margueron, Rudiney Hoffmann~Casali, and Francesca Gulminelli.
\newblock Equation of state for dense nucleonic matter from metamodeling. i.
  foundational aspects.
\newblock {\em Phys. Rev. C}, 97:025805, Feb 2018.

\bibitem{Note1}
Nuclear matter with an equal number of protons and neutrons, i.e. $\delta =0$.

\bibitem{Zdunik2013}
J.~L. {Zdunik} and P.~{Haensel}.
\newblock {Maximum mass of neutron stars and strange neutron-star cores}.
\newblock {\em A\&A}, 551:A61, March 2013.

\bibitem{Alford:2017qgh}
Mark~G. Alford and Armen Sedrakian.
\newblock {Compact stars with sequential QCD phase transitions}.
\newblock {\em Phys. Rev. Lett.}, 119(16):161104, 2017.

\bibitem{Zhang:2023wqj}
Nai-Bo {Zhang} and Bao-An {Li}.
\newblock {Properties of First-Order Hadron-Quark Phase Transition from
  Directly Inverting Neutron Star Observables}.
\newblock {\em arXiv e-prints}, page arXiv:2304.07381, April 2023.

\bibitem{Carreau19}
Thomas Carreau, Francesca Gulminelli, and J\'er\^ome Margueron.
\newblock {Bayesian analysis of the crust-core transition with a compressible
  liquid-drop model}.
\newblock {\em Eur. Phys. J. A}, 55(10):188, 2019.

\bibitem{Ravenhall:1983bdb}
D.~G. Ravenhall, C.~J. Pethick, and J.~M. Lattimer.
\newblock {NUCLEAR INTERFACE ENERGY AT FINITE TEMPERATURES}.
\newblock {\em Nucl. Phys. A}, 407:571--591, 1983.

\bibitem{Maruyama:2005vb}
Toshiki Maruyama, Toshitaka Tatsumi, Dmitri .~N. Voskresensky, Tomonori
  Tanigawa, and Satoshi Chiba.
\newblock {Nuclear pasta structures and the charge screening effect}.
\newblock {\em Phys. Rev. C}, 72:015802, 2005.

\bibitem{Newton:2011dw}
W.~G. Newton, M.~Gearheart, and Bao-An Li.
\newblock {A survey of the parameter space of the compressible liquid drop
  model as applied to the neutron star inner crust}.
\newblock {\em Astrophys. J. Suppl.}, 204:9, 2013.

\bibitem{Wang17}
Meng Wang, G.~Audi, F.~G. Kondev, W.J. Huang, S.~Naimi, and Xing Xu.
\newblock The {AME}2016 atomic mass evaluation ({II}). tables, graphs and
  references.
\newblock {\em Chinese Physics C}, 41(3):030003, mar 2017.

\bibitem{DinhThi21a}
H.~Dinh~Thi, T.~Carreau, A.~F. Fantina, and F.~Gulminelli.
\newblock {Uncertainties in the pasta-phase properties of catalysed neutron
  stars}.
\newblock {\em Astron. Astrophys.}, 654:A114, 2021.

\bibitem{DinhThi21b}
H.~Dinh~Thi, A.~F. Fantina, and F.~Gulminelli.
\newblock {The effect of the energy functional on the pasta-phase properties of
  catalysed neutron stars}.
\newblock {\em Eur. Phys. J. A}, 57(10):296, 2021.

\bibitem{Borhanian21}
S.~{Borhanian}.
\newblock {GWBENCH: a novel Fisher information package for gravitational-wave
  benchmarking}.
\newblock {\em Classical and Quantum Gravity}, 38(17):175014, August 2021.

\bibitem{Cutler:1994ys}
Curt Cutler and Eanna~E. Flanagan.
\newblock {Gravitational waves from merging compact binaries: How accurately
  can one extract the binary's parameters from the inspiral wave form?}
\newblock {\em Phys. Rev. D}, 49:2658--2697, 1994.

\bibitem{Poisson95}
Eric {Poisson} and Clifford~M. {Will}.
\newblock {Gravitational waves from inspiraling compact binaries: Parameter
  estimation using second-post-Newtonian waveforms}.
\newblock {\em Phys. Rev. D}, 52(2):848--855, July 1995.

\bibitem{Valisnieri08}
Michele {Vallisneri}.
\newblock {Use and abuse of the Fisher information matrix in the assessment of
  gravitational-wave parameter-estimation prospects}.
\newblock {\em Phys. Rev. D}, 77(4):042001, February 2008.

\bibitem{Wade14}
Leslie {Wade}, Jolien D.~E. {Creighton}, Evan {Ochsner}, Benjamin~D. {Lackey},
  Benjamin~F. {Farr}, Tyson~B. {Littenberg}, and Vivien {Raymond}.
\newblock {Systematic and statistical errors in a Bayesian approach to the
  estimation of the neutron-star equation of state using advanced gravitational
  wave detectors}.
\newblock {\em \prd}, 89(10):103012, May 2014.

\bibitem{Hinderer08}
Tanja {Hinderer}.
\newblock {Tidal Love Numbers of Neutron Stars}.
\newblock {\em Astrophys. J.}, 677(2):1216--1220, April 2008.

\bibitem{Hinderer09}
Tanja {Hinderer}.
\newblock {Erratum: ``Tidal Love Numbers of Neutron Stars''}.
\newblock {\em Astrophys. J.}, 697(1):964, May 2009.

\bibitem{Pereira20}
Jonas~P. {Pereira}, Micha{\l} {Bejger}, Nils {Andersson}, and Fabian {Gittins}.
\newblock {Tidal Deformations of Hybrid Stars with Sharp Phase Transitions and
  Elastic Crusts}.
\newblock {\em Astrophys. J.}, 895(1):28, May 2020.

\bibitem{Chatziioannou18}
Katerina {Chatziioannou}, Carl-Johan {Haster}, and Aaron {Zimmerman}.
\newblock {Measuring the neutron star tidal deformability with
  equation-of-state-independent relations and gravitational waves}.
\newblock {\em \prd}, 97(10):104036, May 2018.

\bibitem{Dinh-Thi21}
Hoa Dinh~Thi, Chiranjib Mondal, and Francesca Gulminelli.
\newblock The nuclear matter density functional under the nucleonic hypothesis.
\newblock {\em Universe}, 7(10):373, 2021.

\bibitem{Mondal23}
C.~Mondal and F.~Gulminelli.
\newblock Nucleonic metamodeling in light of multimessenger, prex-ii, and crex
  data.
\newblock {\em Phys. Rev. C}, 107:015801, Jan 2023.

\bibitem{Kurkela:2009gj}
Aleksi Kurkela, Paul Romatschke, and Aleksi Vuorinen.
\newblock {Cold Quark Matter}.
\newblock {\em Phys. Rev. D}, 81:105021, 2010.

\bibitem{Xu:2015jwa}
Shu-Sheng Xu, Yan Yan, Zhu-Fang Cui, and Hong-Shi Zong.
\newblock {2+1 flavors QCD equation of state at zero temperature within
  Dyson\textendash{}Schwinger equations}.
\newblock {\em Int. J. Mod. Phys. A}, 30(36):1550217, 2015.

\bibitem{Chen:2015mda}
H.~Chen, J.~B. Wei, M.~Baldo, G.~F. Burgio, and H.~J. Schulze.
\newblock {Hybrid neutron stars with the Dyson-Schwinger quark model and
  various quark-gluon vertices}.
\newblock {\em Phys. Rev. D}, 91(10):105002, 2015.

\bibitem{Zacchi:2015oma}
Andreas Zacchi, Matthias Hanauske, and J\"urgen Schaffner-Bielich.
\newblock {Stable hybrid stars within a SU(3) Quark-Meson-Model}.
\newblock {\em Phys. Rev. D}, 93(6):065011, 2016.

\bibitem{Alvarez-Castillo19}
D.~E. Alvarez-Castillo, D.~B. Blaschke, A.~G. Grunfeld, and V.~P. Pagura.
\newblock Third family of compact stars within a nonlocal chiral quark model
  equation of state.
\newblock {\em Phys. Rev. D}, 99:063010, Mar 2019.

\bibitem{Otto:2019zjy}
Konstantin Otto, Micaela Oertel, and Bernd-Jochen Schaefer.
\newblock {Hybrid and quark star matter based on a nonperturbative equation of
  state}.
\newblock {\em Phys. Rev. D}, 101(10):103021, 2020.

\bibitem{Jokela:2020piw}
Niko Jokela, Matti J\"arvinen, Govert Nijs, and Jere Remes.
\newblock {Unified weak and strong coupling framework for nuclear matter and
  neutron stars}.
\newblock {\em Phys. Rev. D}, 103(8):086004, 2021.

\bibitem{Shahrbaf:2021cjz}
Mahboubeh Shahrbaf, Sofija Anti\'c, A.~Ayriyan, David Blaschke, and
  Ana~Gabriela Grunfeld.
\newblock {Constraining free parameters of a color superconducting nonlocal
  Nambu\textendash{}Jona-Lasinio model using Bayesian analysis of neutron stars
  mass and radius measurements}.
\newblock {\em Phys. Rev. D}, 107(5):054011, 2023.

\bibitem{Drischler16}
C.~Drischler, K.~Hebeler, and A.~Schwenk.
\newblock Asymmetric nuclear matter based on chiral two- and three-nucleon
  interactions.
\newblock {\em Phys. Rev. C}, 93:054314, May 2016.

\bibitem{LIGOScientific:2018hze}
B.~P. Abbott et~al.
\newblock {Properties of the binary neutron star merger GW170817}.
\newblock {\em Phys. Rev. X}, 9(1):011001, 2019.

\bibitem{Ozel:2016oaf}
Feryal \"Ozel and Paulo Freire.
\newblock {Masses, Radii, and the Equation of State of Neutron Stars}.
\newblock {\em Ann. Rev. Astron. Astrophys.}, 54:401--440, 2016.

\bibitem{Kass_bayesfactors_1995}
Robert~E. Kass and Adrian~E. Raftery.
\newblock Bayes factors.
\newblock {\em Journal of the American Statistical Association},
  90(430):773--795, 1995.

\end{thebibliography}
\bibliographystyle{unsrt}




\end{document}